\documentclass[aps,twocolumn,raggedbottom,prb,showpacs,amsmath,amssymb,
superscriptaddress,floatfix]{revtex4}

\usepackage{graphicx}
\usepackage{amscd}
\usepackage{subfigure}

\newcommand{\en}{\epsilon}
\newcommand{\rbr}[1]{\left(#1\right)}
\newcommand{\cbr}[1]{\left[#1\right]}

\newcommand{\pdiff}[2]{\frac{\partial #1}{\partial #2}}

\newcommand{\bv}{\boldsymbol{v}}
\newcommand{\bx}{\boldsymbol{x}}
\newcommand{\bz}{\boldsymbol{z}}
\newcommand{\fh}{\hat{f}}
\newcommand{\gh}{\hat{g}}
\newcommand{\Hh}{\hat{H}}
\newcommand{\tE}{\tau_{\mathrm{E}}}
\newcommand{\tD}{\tau_{\mathrm{d}}}

\newcommand{\rmd}{\mathrm{d}}
\newcommand{\rme}{\mathrm{e}}
\newcommand{\rmi}{\mathrm{i}}

\newcommand{\Tr}{\mathrm{Tr}}

\newcommand{\ie}{\textit{i.e.}\ }

\newcommand{\cf}{\textit{c.f.}\ }

\newcommand{\eref}[1]{(\ref{#1})}
\newcommand{\Eref}[1]{Equation~(\ref{#1})}
\newcommand{\fref}[1]{Fig.~\ref{#1}}
\newcommand{\frefs}[1]{Figs.~\ref{#1}}
\newcommand{\sref}[1]{Sec.~\ref{#1}}
\newcommand{\srefs}[1]{Secs.~\ref{#1}}

\newcommand{\ocite}[1]{Ref.~\onlinecite{#1}}
\newcommand{\ocites}[1]{Refs.~\onlinecite{#1}}

\begin{document}

\title{The Density of States of Chaotic Andreev Billiards}
\author{Jack Kuipers}
\email[]{Jack.Kuipers@physik.uni-regensburg.de}
\author{Thomas Engl}
\email[]{Thomas.Engl@physik.uni-regensburg.de}
\affiliation{Institut f\"ur Theoretische Physik, Universit\"at Regensburg,
D-93040 Regensburg, Germany}
\author{Gregory Berkolaiko}
\affiliation{Department of Mathematics, Texas A\&M University, College Station,
TX 77843-3368, USA}
\author{Cyril Petitjean}
\affiliation{Institut f\"ur Theoretische Physik, Universit\"at Regensburg,
D-93040 Regensburg, Germany}
\affiliation{SPSMS, UMR-E 9001, CEA-INAC/UJF-Grenoble 1, 17 Rue des Martyrs,
38054 Grenoble Cedex 9, France}
\author{Daniel Waltner}
\author{Klaus Richter}
\affiliation{Institut f\"ur Theoretische Physik, Universit\"at Regensburg,
D-93040 Regensburg, Germany}

\date{\today}

\begin{abstract}
Quantum cavities or dots have markedly different properties depending on whether
their classical counterparts are chaotic or not.  Connecting a superconductor to
such a cavity leads to notable proximity effects, particularly the appearance,
predicted by random matrix theory, of a hard gap in the excitation spectrum of
quantum chaotic systems.  Andreev billiards are interesting examples of such
structures built with superconductors connected to a ballistic normal metal
billiard since each time an electron hits the superconducting part it is
retroreflected as a hole (and vice-versa).  Using a semiclassical framework for
systems with chaotic dynamics, we show how this reflection, along with the
interference due to subtle correlations between the classical paths of electrons
and holes inside the system, is ultimately responsible for the gap formation. 
The treatment can be extended to include the effects of a symmetry breaking
magnetic field in the normal part of the billiard or an Andreev billiard
connected to two phase shifted superconductors.   Therefore we are able to see
how these effects can remold and eventually suppress the gap.  Furthermore, the
semiclassical framework is able to cover the effect of a finite Ehrenfest time,
which also causes the gap to shrink.  However for intermediate values this leads
to the appearance of a second hard gap - a clear signature of the Ehrenfest
time.
\end{abstract}


\maketitle

\section{\label{intro}Introduction}

The physics of normal metals (N) in contact with superconductors (S) has been
studied extensively for almost fifty years, and in the past two decades there
has been somewhat of a resurgence of interest in this field.   This has mainly
been sparked by the realization of experiments that can directly probe the
region close to the normal-superconducting (NS) interface at temperatures far
below the transition temperature of the superconductor.  Such experiments have
been possible thanks to microlithographic techniques that permit the building of
heterostructures on a mesoscopic scale combined with transport measurements in
the sub-Kelvin regime.  Such hybrid structures exhibit various new phenomena,
mainly due to the fact that physical  properties of both the superconductor and
the mesoscopic normal metal are strongly influenced by quantum coherence
effects.

The simplest physical picture of this system is that the superconductor tends to
export some of its anomalous properties across the interface over a temperature 
dependent length scale that can be  of the order of a micrometer at low
temperatures.  This is the so-called proximity effect, which has been the focus
on numerous surveys; both
experimental~\cite{gueronetal96,morpurgoetal97,hartogetal97,jakobetal00,mcp01,
vcl01,eromsetal02,escoffieretal04,ecl05} and
theoretical~\cite{lr98,ast00,ta01,beenakker05}.

The key concept to understand this effect~\cite{james64,andreev64,mcmillan68} is
Andreev reflection.  During this process, when an electron from the vicinity of
the Fermi energy ($E_{\rm F}$) surface of the normal conductor hits the
superconductor, the bulk energy gap $\Delta$ of the superconductor prevents the
negative charge from entering, unless a Cooper pair is formed in the
superconductor. Since a Cooper pair is composed of two electrons, an extra
electron has to be taken from the Fermi sea, thus creating a hole in the
conduction band of the normal metal.  Physically and classically speaking, an
Andreev reflection therefore corresponds to a retroflection of the particle,
where Andreev reflected electrons (or holes) retrace their trajectories as holes
(or electrons).  The effect of Andreev reflection on the transport properties of
open NS structures is an interesting and fruitful area (see
\ocites{beenakker97,wt97} and references therein for example), though in this
paper we focus instead on closed structures.  Naturally this choice has the
consequence of leaving aside some exciting recent results such as, for example,
the statistical properties of the conductance~\cite{gma09}, the
magneto-conductance in Andreev quantum dots~\cite{wj09}, resonant
tunneling~\cite{gjw08a}  and the thermoelectrical effect~\cite{cwc09,jw10} in
Andreev interferometers.

In closed systems, one of the most noticeable manifestations of the proximity
effect is the suppression of the density of states (DoS) of the normal metal
just above the Fermi energy.  Although most of the experimental investigations
have been carried out on disordered
systems~\cite{gueronetal96,hartogetal97,mcp01,vcl01,escoffieretal04}, with
recent technical advances interest has moved to structures with clean ballistic
dynamics~\cite{morpurgoetal97,jakobetal00,eromsetal02,ecl05,choietal05,ew07}. 
This shift  gives access to the experimental investigation of the so-called
\emph{Andreev billiard}.  While this term was originally coined~\cite{kmg95} for
an impurity-free normal conducting region entirely confined by a superconducting
boundary, it also refers to a ballistic normal area (\ie a quantum dot) with a
boundary that is only partly connected to a superconductor.  The considerable
theoretical attention raised by such a hybrid structure in the past decade is
related to the interesting peculiarity that by looking at the DoS of an Andreev
billiard we can determine the nature of the underlying dynamics of its classical
counterpart~\cite{melsenetal96}.  Indeed, while the DoS vanishes with a power
law in energy for the integrable case, the spectrum of a chaotic billiard is
expected to exhibit a true gap above $E_{\rm F}$~\cite{melsenetal96}.  The width
of this hard gap, also called the minigap~\cite{beenakker05}, has been
calculated as a purely quantum effect by using random matrix theory (RMT) and
its value scales with the Thouless energy, $E_{\rm T} = \hbar /2\tD$, where
$\tD$ is the average (classical) dwell time a particle stays in the billiard
between successive Andreev  reflections~\cite{melsenetal96}.

Since the existence of this gap is expected to be related to the chaotic nature
of the electronic motion, many attempts have been undertaken to explain this
result in semiclassical
terms~\cite{ln98,sb99,ihraetal01,ir01,csertietal02,ag02,zaitsev06},  however
this appeared to be rather complicated.  Indeed a traditional semiclassical
treatment based on the so-called Bohr-Sommerfeld  (BS) approximation yields only
an exponential suppression of the DoS~\cite{ln98,sb99,ihraetal01}.  This
apparent contradiction of this prediction with the RMT one was resolved quite
early by Lodder and Nazarov~\cite{ln98} who pointed out the existence of two
different regimes.  The characteristic time scale that governs  the crossover
between the two regimes is the Ehrenfest time $\tE \sim \vert\ln\hbar\vert$,
which is the time scale that separates the evolution of wave packets following
essentially the classical dynamics from longer time scales dominated by wave
interference.  In particular it is the ratio $\tau=\tE/\tD$,  that has to be
considered.  

In the universal regime,  $\tau = 0$, chaos sets in sufficiently rapidly and RMT
is valid leading to the appearance of the aforementioned Thouless
gap~\cite{melsenetal96}.  Although the Thouless energy $E_{\rm T}$ is related to
a purely classical quantity, namely the average dwell time, we stress that the
appearance of the minigap is a quantum mechanical effect, and consequently the
gap closes if a symmetry breaking magnetic field is applied~\cite{melsenetal97}.
 Similarly if two superconductors are attached to the Andreev billiard, the size
of the gap will depend on the relative phase between the two superconductors,
with the gap vanishing for a $\pi$-junction~\cite{melsenetal97}.

The deep classical limit is characterized by $\tau\to\infty$, and in this regime
the suppression of the DoS is exponential and well described by the
BS approximation. The more interesting crossover regime of finite
Ehrenfest time, and the conjectured Ehrenfest time gap dependence of
\ocite{ln98} have been investigated by various
means~\cite{ta01,ab02,sgb03,vl03,br06b,gjw08a,ma09}.  Due to the logarithmic
nature of $\tE$, investigating numerically the limit of large Ehrenfest time is
rather difficult, but a clear signature of the gap's Ehrenfest time dependence
has been obtained~\cite{jsb03,kormanyosetal04,sj05} for $\tau<1$.  From an
analytical point of view RMT is inapplicable in the finite $\tE$
regime~\cite{ta01}, therefore new methods such as a stochastic
method~\cite{vl03} using smooth disorder and sophisticated perturbation methods
that include diffraction effects~\cite{ab02} have been used to tackle this
problem.  On the other hand a purely phenomenological model, effective RMT, has
been developed~\cite{sgb03, gjb05} and predicts a gap size scaling with the
Ehrenfest energy  $E_{\rm E} = \hbar/2\tE$.  Recently Micklitz and
Altland~\cite{ma09}, based on a refinement of the quasiclassical approach and
the Eilenberger equation, succeeded to show the existence of a gap of width $\pi
E_{\rm E} \propto 1/\tau$ in the limit of large $\tau\!\gg\!1$.

Consequently a complete picture of all the available regimes was still missing
until recently when we treated the DoS semiclassically~\cite{kuipersetal10}
following the scattering approach~\cite{beenakker91}.  Starting for $\tau = 0$
and going beyond the diagonal approximation we used an energy-dependent
generalization of the work~\cite{bhn08} on the moments of the transmission
eigenvalues. The calculation is based on the evaluation of correlation functions
also appearing in the moments of the Wigner delay times~\cite{bk10}.  More
importantly, the effect of finite Ehrenfest time could be incorporated in this
framework~\cite{waltneretal10} leading to a microscopic confirmation of the
$\tE$ dependence of the gap predicted by effective RMT.  Interestingly the
transition between $\tau=0$ and $\tau=\infty$ is not smooth and a second gap at
$\pi E_{\rm E}$ was observed for intermediate $\tau$, providing us with
certainly the most clear-cut signature of Ehrenfest time effects.

In this paper we extend and detail the results obtained
in~\cite{kuipersetal10}.  First we discuss Andreev billiards and their treatment
using RMT and semiclassical techniques.  For the DoS in the universal regime
($\tau=0$) we first delve into the work of \ocites{bhn08,bk10} before using it
to obtain the generating function of the correlation functions which are
employed to derive the DoS. This is done both in the absence and in the presence
of a time reversal symmetry breaking magnetic field, and we also look at the
case when the bulk superconducting gap and the excitation energy of the particle
are comparable.

We then treat Andreev billiards connected to two superconducting contacts with a
phase difference $\phi$.  The gap is shown to shrink with increasing phase
difference due to the the accumulation of a phase along the trajectories that
connect the two superconductors.  Finally the Ehrenfest regime will be
discussed, especially the appearance of a second intermediate gap for a certain
range of $\tau$.  We will also show that this intermediate gap is very sensitive
to the phase difference between the superconductors.

\section{Andreev billiards}

\begin{figure}
\includegraphics[width=0.85\columnwidth]{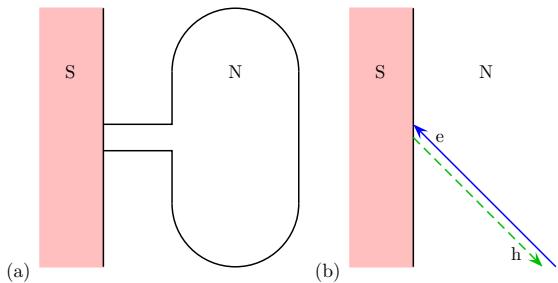}
\caption{\label{andreevbilliard}(a) The Andreev billiard consists of a chaotic
normal metal (N) cavity attached to a superconductor (S) via a lead.  (b) At the
NS interface between the normal metal and the superconductor electrons are
retroreflected as holes.}
\end{figure}

Since the treatment of Andreev billiards was recently reviewed in
\ocite{beenakker05} we just recall some useful details here.  In particular the
chaotic Andreev billiard that we consider is treated within the scattering
approach~\cite{beenakker91} where the NS interface is modelled with the help of
a fictitious ideal lead.  This lead permits the contact between the normal metal
cavity (with chaotic classical dynamics) and the semi-infinite superconductor as
depicted in \fref{andreevbilliard}a.

Using the continuity of the superconducting and normal wave function, we can
construct the scattering matrix of the whole system.  Denoting the excitation
energy of the electron above the Fermi energy $E_{\mathrm{F}}$ by $E$ and
assuming that the lead supports $N$ channels (transverse modes at the Fermi
energy), the scattering matrix of the whole normal region can be written in a
joint electron-hole basis and reads 
\begin{equation}
S_{\mathrm{N}}(E)=\left(\begin{array}{cc}
S(E) & 0 \\
0 & S^{*}(-E) 
\end{array}\right) ,
\end{equation}
where $S(E)$ is the unitary $N\times N$ scattering matrix of the electrons (and
its complex conjugate $S^{*}(-E)$ that of the holes).  As the electrons and
holes remain uncoupled in the normal region the off-diagonal blocks are zero. 
Instead, electrons and holes couple at the NS interface through Andreev
reflection~\cite{andreev64} where electrons are retroreflected as holes and
vice versa, as in \fref{andreevbilliard}b. For energies $E$ smaller than the
bulk superconductor gap $\Delta$ there is no propagation into the superconductor
and if we additionally assume $\Delta\ll E_{\mathrm{F}}$ we can encode the
Andreev reflection in the matrix
\begin{equation}
S_{\mathrm{A}}(E)=\alpha(E)\left(\begin{array}{cc}
0 & 1 \\
1 & 0 
\end{array}\right) ,
\end{equation}
\begin{equation}
\alpha(E) = \rme^{-\rmi \arccos\left(\frac{E}{\Delta}\right)} =
\frac{E}{\Delta}-\rmi\sqrt{1-\frac{E^2}{\Delta^2}} .
\label{alphadefeqn}
\end{equation}
The retroreflection (of electrons as holes with the same channel index) is
accompanied by the phase shift $\arccos\left(E/\Delta\right)$.  In the limit of
perfect Andreev reflection ($E=0$) this phase shift reduces to $\pi/2$.

Below $\Delta$ the Andreev billiard has a discrete excitation spectrum at
energies where $\det\cbr{1-S_{\mathrm{A}}(E)S_{\mathrm{N}}(E)}=0$, which can be
simplified~\cite{beenakker91} to
\begin{equation}
\label{determinantaleq}
\det\cbr{1-\alpha^2(E)S(E)S^{*}(-E)}=0 .
\end{equation}
Finding the roots of this equation yields the typical density of states of
chaotic Andreev billiards.  In the next two Sections we review the two main
analytical frameworks that can be used to tackle this problem.

\subsection{\label{rmt}Random matrix theory}

\begin{figure}
\includegraphics[width=1\columnwidth]{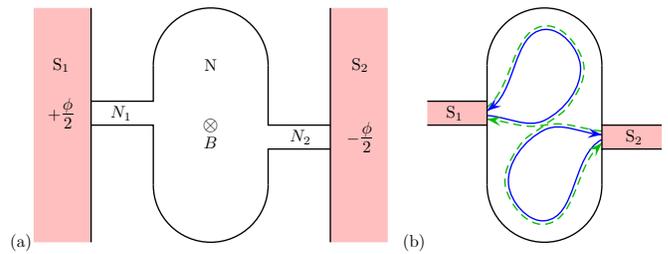}
\caption{\label{andreevgeneralbilliard} (a) An Andreev billiard connected to two
superconductors (S$_1$, S$_2$) at phases $\pm\phi/2$ via leads carrying $N_{1}$
and $N_{2}$ channels, all threaded by a perpendicular magnetic field $B$. (b)
The semiclassical treatment involves classical trajectories retroreflected at
the superconductors an arbitrary number of times.}
\end{figure}

One powerful treatment uses random matrix theory.  Such an approach was
initially considered in \ocites{melsenetal96,melsenetal97} where the actual
setup treated is depicted in \fref{andreevgeneralbilliard}a.  It consists of a
normal metal (N) connected to two superconductors (S$_1$, S$_2$) by narrow leads
carrying $N_{1}$ and $N_{2}$ channels.  The superconductors' order parameters
are considered to have phases $\pm\phi/2$, with a total phase difference $\phi$.
Moreover a perpendicular magnetic field $B$ was applied to the normal part.  We
note that although this figure (and \fref{andreevbilliard}a) have spatial
symmetry the treatment is actually for the case without such symmetry.

As above, the limit $\Delta\ll E_\mathrm{F}$ was taken so that normal reflection
at the NS interface can be neglected and the symmetric case in which both leads
contain the same number, $N/2$, of channels was
considered~\cite{melsenetal96,melsenetal97}.  Finally it was also assumed that
$\alpha\approx-\rmi$, valid in the limit $E,E_\mathrm{T}\ll\Delta\ll
E_\mathrm{F}$. For such a setup, the determinantal equation
\eref{determinantaleq} becomes
\begin{equation}
 \det\cbr{1+S(E)\rme^{\rmi\tilde{\phi}}S^{*}(-E)\rme^{-\rmi\tilde{\phi}}}=0 ,
 \label{deteqdos}
\end{equation}
where $\tilde{\phi}$ is a diagonal matrix whose first $N/2$ elements are
$\phi/2$ and the remaining $N/2$ elements $-\phi/2$.  We note that though we
stick to the case of perfect coupling here, the effect of tunnel barriers was
also included in \ocite{melsenetal96}.

The first step is to rewrite the scattering problem in terms of a low energy
effective Hamiltonian ${\cal H}$
\begin{equation}
{\cal H} = 
\left(
\begin{array}{cc}
\Hh & \pi XX^{T}\\
-\pi XX^{T} & -\Hh^{*}
\end{array}
\right) ,
\end{equation}
where $\Hh$ is the $M\times M$ Hamiltonian of the isolated billiard and $X$ an
$M\times N$ coupling matrix.  Eventually the limit $M\to\infty$ is taken and to
mimic a chaotic system the matrix $\Hh$ is replaced by a random matrix following
the Pandey-Mehta distribution~\cite{beenakker97}
\begin{eqnarray}\label{pandeymehta}
 P(H)&\propto&\exp\left(-\frac{N^2\rbr{1+a^2}}{64ME_{\mathrm{T}}^2}\right. \\
&& \left. \qquad \times \sum_{i,j=1}^{M}
\cbr{\rbr{\mathrm{Re}\Hh_{ij}}^2+a^{-2}\rbr{\mathrm{Im}\Hh_{ij}}^2}\right) .
\nonumber 
\end{eqnarray}
The parameter $a$ measures the strength of the time-reversal symmetry breaking
so we can investigate the crossover from the ensemble with time-reversal
symmetry, the Gaussian orthogonal ensemble (GOE), to that without, the Gaussian
unitary ensemble (GUE).  It is related to the magnetic flux $\Phi$ through the
two-dimensional billiard of area A and with Fermi velocity $v_{\mathrm{F}}$ by
\begin{equation}
 Ma^2=c\rbr{\frac{e\Phi}{h}}^2\hbar v_{\mathrm{F}}\frac{N}{2\pi
E_{\mathrm{T}}\sqrt{A}} .
 \label{trbreaking}
\end{equation}
Here $c$ is a numerical constant of order unity depending only on the shape of
the billiard. The critical flux is then defined via
\begin{equation}
 Ma^2=\frac{N}{8}\rbr{\frac{\Phi}{\Phi_{\mathrm{c}}}}
^2\qquad\Leftrightarrow\qquad\Phi_{\mathrm{c}}\approx\frac{h}{e}\rbr{\frac{2\pi
E_{\mathrm{T}}}{\hbar v_{\mathrm{F}}}}^\frac{1}{2}A^\frac{1}{4} .
 \label{critflux}
\end{equation}

The density of states, divided for convenience by twice the mean density of
states of the isolated billiard, can be written as
\begin{equation}
 d(\en)=-\mathrm{Im}W(\en) ,
 \label{rmtdoseq}
\end{equation}
where $W(\en)$ is the trace of a block of the Green function of the effective
Hamiltonian of the scattering system and for simplicity here we express the
energy in units of the Thouless energy $\en=E/E_{\mathrm{T}}$.  This is averaged
by integrating over \eref{pandeymehta} using diagrammatic methods~\cite{bb96},
which to leading order in inverse channel number $1/N$ leads to the
expression~\cite{melsenetal97}
\begin{equation}
 W(\en)=\rbr{\frac{b}{2}W(\en)-\frac{\en}{2}}\rbr{1+W^{2}(\en)+\frac{\sqrt{1+W^{
2}(\en)}}{\beta}} ,
 \label{origrmteq}
\end{equation}
where $\beta=\cos\rbr{\phi/2}$ and $b=\rbr{\Phi/\Phi_\mathrm{c}}^{2}$ with the
critical magnetic flux $\Phi_\mathrm{c}$ for which the gap in the density of
states closes (at $\phi=0$). \Eref{origrmteq} may also be rewritten as a sixth
order polynomial and when substituting into \eref{rmtdoseq}, we should take the
solution that tends to 1 for large energies.  In particular, when there is no
phase difference between the two leads ($\phi=0$, or equivalently when we
consider a single lead carrying $N$ channels) and no magnetic field in the
cavity ($\Phi/\Phi_\mathrm{c}=0$) the density of states is given by a solution
of the cubic equation
\begin{equation}
\epsilon^2W^3(\en)+4\en W^2(\en)+(4+\en^2)W(\en)+4\en=0 .
 \label{rmtsimplest}
\end{equation}

\subsection{\label{semiapproach}Semiclassical approach}

The second approach, and that which we pursue and detail in this paper, is to
use the semiclassical approximation to the scattering matrix which involves the
classical trajectories that enter and leave the cavity~\cite{miller75}.  Using
the general expression between the density of states and the scattering
matrix~\cite{akkermansetal91}, the density of states of an Andreev billiard
reads~\cite{beenakker91,ds92,ihraetal01}
\begin{equation}
\label{semidosstarteq}
\tilde{d}(E)=\bar{d} - \frac{1}{\pi}\mathrm{Im} \pdiff{}{E} \ln \det
\cbr{1-S_{\mathrm{A}}(E)S_{\mathrm{N}}(E)} ,
\end{equation}
where $\bar{d}=N/2\pi E_{\mathrm{T}}$ is twice the mean density of states of the
isolated billiard (around the Fermi energy).  \Eref{semidosstarteq} should be
understood as an averaged quantity over a small range of the Fermi energy or
slight variations of the billiard  and for convergence reasons a small imaginary
part is included in the energy $E$.  In the limit of perfect Andreev reflection
$\alpha(E)\approx-\rmi$, see \eref{alphadefeqn}, and \eref{semidosstarteq}
reduces to
\begin{equation}
\tilde{d}(E)=\bar{d} + \frac{1}{\pi}\mathrm{Im} \pdiff{}{E} \Tr
\sum_{m=1}^{\infty} \frac{1}{m} \left(\begin{array}{cc}
0 & \rmi S^{*}(-E) \\
\rmi S(E)  & 0
\end{array}\right)^{m} .
\end{equation}
Obviously only even terms in the sum have a non-zero trace, and setting $n=2m$,
dividing through by $\bar{d}$ and expressing the energy in units of the Thouless
energy $\en=E/E_{\mathrm{T}}$, this simplifies to~\cite{ihraetal01}
\begin{equation}
d(\en)=1 + 2 \mathrm{Im} \sum_{n=1}^{\infty} \frac{(-1)^{n}}{n}
\pdiff{C(\en,n)}{\en} .
\label{dossemieqn}
\end{equation}
\Eref{dossemieqn} involves the correlation functions of $n$ scattering matrices
\begin{equation}
C(\en,n)=\frac{1}{N}\Tr
\cbr{S^{*}\rbr{-\frac{\en\hbar}{2\tD}}S\rbr{\frac{\en\hbar}{2\tD}}}^n ,
 \label{singlecorr}
\end{equation}
where we recall that the energy is measured relative to the Fermi energy and
that $E_{\mathrm{T}}=\hbar/2\tD$ involves the average dwell time $\tD$.   For
chaotic systems~\cite{lv04b} the dwell time can be expressed as
$\tD=T_{\mathrm{H}}/N$ in terms of the Heisenberg time $T_{\mathrm{H}}$
conjugate to the mean level spacing ($2/\bar{d}$).

At this point it is important to observe that nonzero values of $\en$ are
necessary for the convergence of the expansion of the logarithm in
\eref{semidosstarteq} that led to \eref{dossemieqn}.  On the other hand, we are
particularly interested in small values of $\en$ which puts \eref{dossemieqn} on
the edge of the radius of convergence, where it is highly oscillatory.  The
oscillatory behavior and a slow decay in $n$ is a direct consequence of the
unitarity of the scattering matrix at $\en=0$ (in fact later it can also be
shown that $\pdiff{C(\en,n)}{\en}\vert_{\en=0}=\rmi n$).  Thus a truncation of
\eref{dossemieqn} will differ markedly from the predicted RMT gap, which was the
root of the difficulty of capturing the gap by previous semiclassical
treatments~\cite{ihraetal01,ag02,zaitsev06}.  In the present work we succeed in
evaluating the entire sum and hence obtain results which are uniformly valid for
all values of $\en$.

Calculating the density of states is then reduced to the seemingly more
complicated task of evaluating correlation functions semiclassically for all
$n$.  Luckily the treatment of such functions has advanced rapidly in the last
few years~\cite{rs02,heusleretal06,mulleretal07,bhn08,bk10} and we build on that
solid basis.  We also note that determining $C(\en,n)$ is a more general task
than calculating the density of states.  Since the Andreev reflection has
already been encoded in the formalism before \eref{dossemieqn}, the treatment of
the $C(\en,n)$ no longer depends on the presence or absence of the
superconducting material, but solely on the properties of the chaotic dynamics
inside the normal metal billiard.
\begin{figure*}
\includegraphics[width=0.85\textwidth]{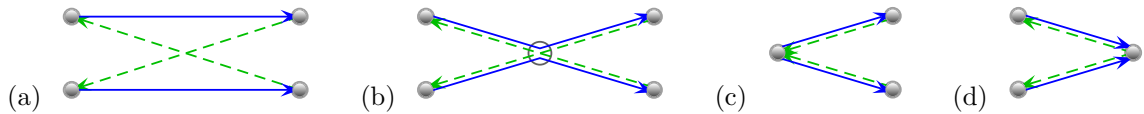}
\caption{\label{twotrajectorystructures}(a) The original trajectory structure of
the correlation function $C(\en,2)$ where the incoming channels are drawn on the
left, outgoing channels on the right, electrons as solid (blue) and holes as
dashed (green) lines. (b) By collapsing the electron trajectories directly onto
the hole trajectories we create a structure where the trajectories only differ
in a small region called an encounter.  Placed inside the Andreev billiard this
diagram corresponds to \fref{andreevgeneralbilliard}b.  The encounter can be
slid into the incoming channels on the left (c) or the outgoing channels on the
right (d) to create diagonal-type pairs.}
\end{figure*}

In the semiclassical approximation, the elements of the scattering matrix are
given by~\cite{miller75}
\begin{equation}
 S_{o i}(E)\approx\frac{1}{\sqrt{T_\mathrm{H}}}\sum_{\zeta(i\rightarrow
o)}A_\zeta\rme^{\rmi S_{\zeta}(E)/\hbar} ,
 \label{semiclscat}
\end{equation}
where the sum runs over all classical trajectories $\zeta$ starting in channel
$i$ and ending in channel $o$. $S_\zeta(E)$ is the classical action of the
trajectory $\zeta$ at energy $E$ above the Fermi energy and the amplitude
$A_\zeta$ contains the stability of the trajectory as well as the Maslov
phases~\cite{richter00}.  After we substitute \eref{semiclscat} into
\eref{singlecorr} and expand the action around the Fermi energy up to first
order in $\en$ using $\partial S_\zeta/\partial E=T_\zeta$ where $T_\zeta$ is
the duration of the trajectory $\zeta$, the correlation functions are given
semiclassically by a sum over $2n$ trajectories
\begin{eqnarray} 
C(\en,n)&\approx&\frac{1}{NT_{\mathrm{H}}^n}\prod_{j=1}^{n}\sum_{i_j,o_j}\sum_{
\substack{\zeta_j(i_{j}\rightarrow o_{j})\cr \zeta'_j(o_{j}\rightarrow
i_{j+1})}}A_{\zeta_j}
A_{\zeta'_j}^*\rme^{\rmi(S_{\zeta_j}-S_{\zeta'_j})/\hbar}\nonumber \\
& & \quad \times \rme^{\rmi\en(T_{\zeta_j}+T_{\zeta'_j})/(2\tD)} .
\label{singlesemiclcorr}
\end{eqnarray}
The final trace in \eref{singlecorr} means that we identify $i_{n+1}=i_1$ and as
the electron trajectories $\zeta_j$ start at channel $i_j$ and end in channel
$o_j$ while the primed hole trajectories $\zeta'_j$ go backwards starting in
channel $o_j$ and ending in channel $i_{j+1}$ the trajectories fulfill a complete
cycle, as in \frefs{twotrajectorystructures}a
and~\ref{fourtrajectorystructures}a,d,g. The channels $i_1,\ldots,i_n$ will be
referred to as incoming channels, while $o_1,\ldots,o_n$ will be called outgoing
channels.  This refers to the direction of the electron trajectories at the
channels and not necessarily to which lead the channel finds itself in (when we
have two leads as in \fref{andreevgeneralbilliard}).

The actions in \eref{singlesemiclcorr} are taken at the Fermi energy and the
resulting phase is given by the difference of the sum of the actions of the
unprimed trajectories and the sum of the actions of the primed ones.   In the
semiclassical limit of $\hbar\to0$ (\cf the RMT limit of $M\to\infty$) this
phase oscillates widely leading to cancellations when the averaging is applied,
unless this total action difference is of the order of $\hbar$.  The
semiclassical treatment then involves finding sets of classical trajectories
that can have such a small action difference and hence contribute consistently
in the limit $\hbar\to0$.

\section{Semiclassical diagrams}

\begin{figure*}
\includegraphics[width=0.85\textwidth]{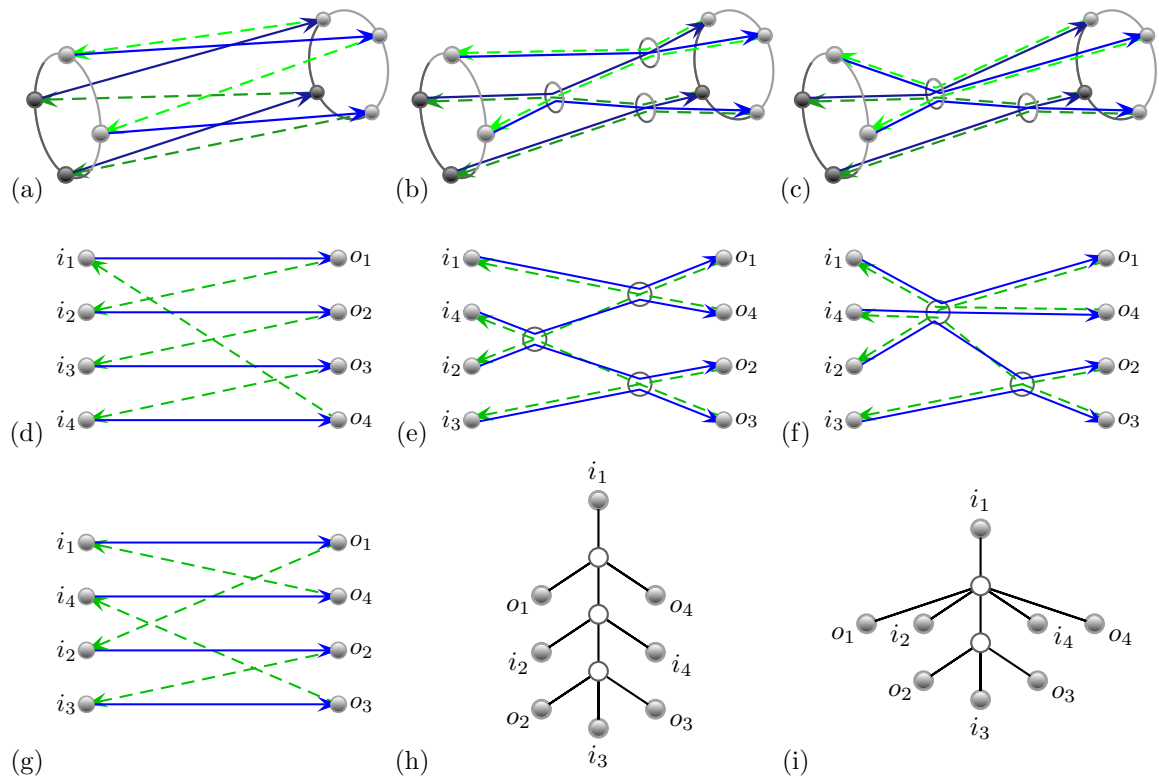}
\caption{\label{fourtrajectorystructures}(a) The original trajectory structure
of the correlation function $C(\en,4)$ where the incoming channels are drawn on
the left, outgoing channels on the right, electrons as solid (blue) and holes as
dashed (green) lines. (d,g) Equivalent 2D projections of the starting structure
as the order is determined by moving along the closed cycle of electron and hole
trajectories. (b) By pinching together the electron trajectories (pairwise here)
we can create a structure which only differs in three small regions (encounters)
and which can have a small action difference.  (e) Projection of (b) also
created by collapsing the electron trajectories in (g) directly onto the hole
trajectories.  (c,f) Sliding two of the encounters from (b) together (or
originally pinching 3 electron trajectories together) creates these diagrams. 
(h,i) Resulting rooted plane tree diagrams of (e,f) or (b,c) defining the top
left as the first incoming channel (\ie the channel ordering as depicted in
(e,f)).}
\end{figure*}

As an example we show the original trajectory structure for $n=2$ in
\fref{twotrajectorystructures}a, where for convenience we draw the incoming
channels on the left and the outgoing channels on the right so that electrons
travel to the right and holes to the left (\cf the shot noise in
\ocites{braunetal06,wj06,lassl03}).  Of course the channels are really in the
lead (\fref{andreevbilliard}a) or either lead (\fref{andreevgeneralbilliard})
and the trajectory stretches involve many bounces at the normal boundary of the
cavity.  We draw such topological sketches as the semiclassical methods were
first developed for transport~\cite{rs02,mulleretal07,bhn08} where typically we
have $S^{\dagger}$ (complex conjugate transpose) instead of $S^{*}$ (complex
conjugate) in \eref{singlecorr}, restricted to the transmission subblocks, so
that all the trajectories would travel to the right in our sketches.  Without
the magnetic field, the billiard has time reversal symmetry and $S$ is
symmetric, but this difference plays a role when we turn the magnetic field on
later.  An even more important difference is that in our problem any channel can
be in any lead.

To obtain a small action difference, and a possible contribution in the
semiclassical limit, the trajectories must be almost identical.  This can be
achieved for example by collapsing the electron trajectories directly onto the
hole trajectories as in \fref{twotrajectorystructures}b.  Inside the open
circle, the holes still `cross' while the electrons `avoid crossing', but by
bringing the electron trajectories arbitrarily close together the set of
trajectories can have an arbitrary small action difference.  More accurately,
the existence of partner trajectories follows from the hyperbolicity of the
phase space dynamics.  Namely, given two electron trajectories that come close
(have an encounter) in the phase space, one uses the local stable and unstable
manifolds~\cite{sr01,spehner03,tr03} to find the coordinates through which hole
trajectories arrive along one electron trajectory and leave along the other,
exactly as in \fref{twotrajectorystructures}b (and
\fref{andreevgeneralbilliard}b).  These are the partner trajectories we pick for
$\zeta'_1$ and $\zeta'_2$ when we evaluate $C(\en,2)$ from
\eref{singlesemiclcorr} in the semiclassical approximation.  As the encounter
involves two electron trajectories it is called a 2-encounter.  An encounter can
happen anywhere along the length of a trajectory.  In particular, it can happen
at the very beginning or the very end of a trajectory, in which case it is
actually happening next to the lead, see \frefs{twotrajectorystructures}c,d. 
This situation is important as it will give an additional contribution to that of
an encounter happening in the body of the billiard.  We will refer to this
situation as an `encounter entering the lead'.  We note that if an encounter
enters the lead the corresponding channels must coincide and we have
diagonal-type pairs (\ie the trajectories are coupled exactly pairwise) though
it is worth bearing in mind that there is still a partial encounter happening
near the lead as shown by the Ehrenfest time treatment~\cite{wj06,br06}.

To give a more representative example, consider the structure of trajectories
for $n=4$.  For visualization purposes in \fref{fourtrajectorystructures}a the
original trajectories are arranged around a cylinder in the form of a cat's
cradle.  The incoming and outgoing channels are ordered around the circles at
either end although they could physically be anywhere.  Projecting the structure
into 2D we can draw it in several equivalent ways, for example as in
\fref{fourtrajectorystructures}d or~\ref{fourtrajectorystructures}g, and we must
take care not to overcount such equivalent representations.  We note that the
ordering of the channels is uniquely defined by the closed cycle that the
trajectories form.  To create a small action difference, we can imagine pinching
together the electron (and hole) strings in \fref{fourtrajectorystructures}a. 
One possibility is to pinch two together in three places (making three
2-encounters) as in \fref{fourtrajectorystructures}b.  A possible representation
in 2D is shown in \fref{fourtrajectorystructures}e, which can also be created by
collapsing the electron trajectories directly onto the hole trajectories in
\fref{fourtrajectorystructures}g.  Note that the collapse of the diagram in
\fref{fourtrajectorystructures}d leads to a different structure with three
2-encounters.  However in general it is not true that the different projections
of the arrangement in \fref{fourtrajectorystructures}a are in a one-to-one
correspondence with all possible diagrams.

\begin{figure*}
\includegraphics[width=0.85\textwidth]{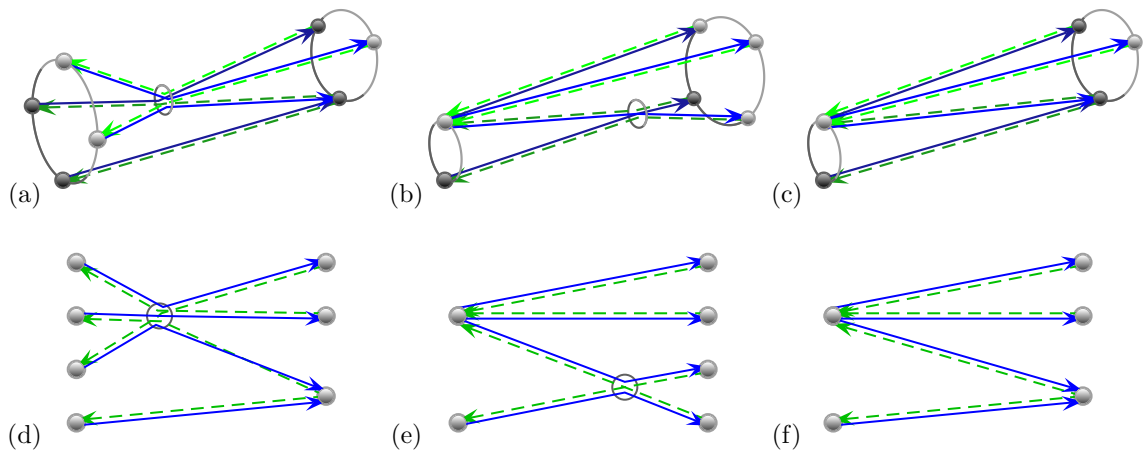}
\caption{\label{fourtrajectorystructureleads} Further possibilities arise from
moving encounters into the lead(s). Starting from
\fref{fourtrajectorystructures}c we can slide the 2-encounter into the outgoing
channels on the right (called `$o$-touching', see text) to arrive at (a,d) or
the 3-encounter into the incoming channels on the left (called `$i$-touching')
to obtain (b,e).  Moving both encounters leads to (c,f), but moving both to the
same side means first combining the 3- and 2-encounter in
\fref{fourtrajectorystructures}c into a 4-encounter and is treated as such.}
\end{figure*}

From \frefs{fourtrajectorystructures}b,e we can create another possibility by
sliding two of the 2-encounters together to make a 3-encounter (or alternatively
we could start by pinching three trajectories together in
\fref{fourtrajectorystructures}a as well as an additional pair) as in
\fref{fourtrajectorystructures}c,f.  Finally we could combine both to a single
4-encounter.  Along with the possibilities where all the encounters are inside
the system, we can progressively slide encounters into the leads, as we did for
the $n=2$ case in \fref{twotrajectorystructures}, creating, among others, the
diagrams in \fref{fourtrajectorystructureleads}.

Finally, we mention that so far we were listing only `minimal' diagrams.  One
can add more encounters to the above diagrams but we will see later that such
arrangements contribute at a higher order in the inverse number of channels and
are therefore subdominant.  The complete expansion in this small parameter is
available only for small values of $n$, see
\ocites{heusleretal06,braunetal06,mulleretal07}.

\subsection{\label{treerecursions}Tree recursions}

To summarize the previous paragraph, the key task now is to generate all
possible minimal encounter arrangements (see, for example, \ocite{bk10} for the
complete list of those with $n=3$).
This is a question that was answered in \ocite{bhn08} where the moments of the
transmission amplitudes were considered.  The pivotal step was to redraw the
diagrams as rooted plane trees and to show that there is a one-to-one relation
between them (for the diagrams that contribute at leading order in inverse
channel number).  To redraw a diagram as a tree we start with a particular
incoming channel $i_{1}$ as the root (hence rooted trees) and place the
remaining channels in order around an anticlockwise loop (hence plane).  Moving
along the trajectory $\zeta_{1}$ we draw each stretch as a link and each
encounter as a node (open circle) until we reach $o_1$.  Then we move along
$\zeta'_1$ back to its first encounter and continue along any new encounters to
$i_2$ and so on.  For example, the tree corresponding to
\frefs{fourtrajectorystructures}b,e is drawn in \fref{fourtrajectorystructures}h
and that corresponding to \frefs{fourtrajectorystructures}c,f is in
\fref{fourtrajectorystructures}i.  Note that marking the root only serves to
eliminate overcounting and the final results do not depend on the particular
choice of the root.

A particularly important property of the trees is their amenability to recursive
counting.  The recursions behind our treatment of Andreev billiards were derived
in \ocite{bhn08} and we recall the main details here.  First we can describe the
encounters in a particular tree by a vector $\bv$ whose elements $v_l$ count the
number of $l$-encounters in the tree (or diagram); this is often written as
$2^{v_2}3^{v_3}\cdots$.  An $l$-encounter is a vertex in the tree of degree $2l$
(\ie connected to $2l$ links).  The vertices of the tree that correspond to
encounters will be called `nodes', to distinguish them from the vertices of
degree 1 which correspond to the incoming and outgoing channels and which will
be called `leaves'.  The total number of nodes is $V=\sum_{l>1}v_l$ and the
number of leaves is $2n$ where $n$ is the order of the correlation function
$C(\en,n)$ to which the trees contribute.  Defining $L=\sum_{l>1}lv_l$, we can
express $n$ as $n=(L-V+1)$.  Note that the total number of links is $L+n$ which
can be seen as $l$ links trailing each $l$-encounter plus another $n$ from the
incoming channels.  For example, the $2^{1}3^{1}$ tree in
\fref{fourtrajectorystructures}i has $L=5$, $V=2$ and contributes to the $n=4$
correlation function.  We always draw the tree with the leaves ordered
$i_{1},o_{1},\ldots,i_{n},o_{n}$ in anticlockwise direction.  This fixes the
layout of the tree in the plane, thus the name `rooted plane
trees'~\cite{tutte64}.

From the start tree, we can also move some encounters into the lead(s) and it is
easy to read off when this is possible.  If an $l$-encounter (node of degree
$2l$) is adjacent to exactly $l$ leaves with label $i$ it may `$i$-touch' the
lead, \ie the electron trajectories have an encounter upon entering the system
and the corresponding incoming channels coincide.  Likewise if a $2l$-node is
adjacent to $l$ $o$-leaves it may `$o$-touch' the lead.  For example, in
\fref{fourtrajectorystructures}i the top node has degree 6, is adjacent to 3
$i$-leaves (including the root) and can $i$-touch the lead as in
\frefs{fourtrajectorystructureleads}b,e.  The lower encounter can $o$-touch as
in \frefs{fourtrajectorystructureleads}a,d.  In addition, both encounters can
touch the lead to create \frefs{fourtrajectorystructureleads}c,f.
\begin{figure*}
\includegraphics[width=0.75\textwidth]{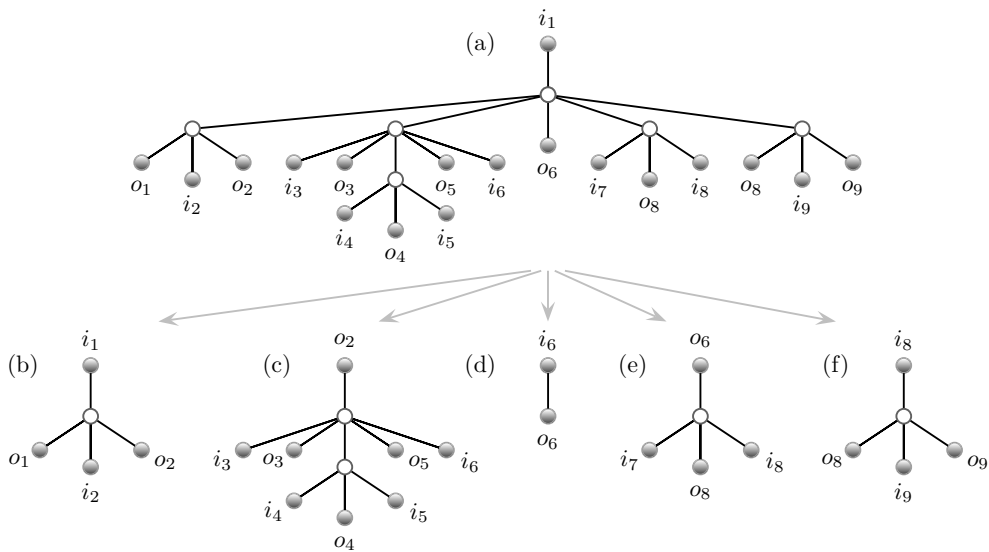}
\caption{\label{treecutting}The tree shown in (a) is cut at its top node (of
degree 6) such that the trees (b)-(f) are created.  Note that to complete the
five new trees we need to add an additional four new links and leaves and that
the trees (c) and (e) in the even positions have the incoming and outgoing
channels reversed.}
\end{figure*}

Semiclassically, we add the contributions of all the possible trajectory
structures (or trees) and the contribution of each is made up by multiplying the
contributions of its constituent parts (links, encounters and leaves).  First we
count the orders of the number of channels $N$.  As mentioned in \ocite{bhn08}
(see also \sref{singlelead} below) the multiplicative contribution of each
encounter or leaf is of order $N$ and each link gives a contribution of order
$1/N$.  Together with the overall factor of $1/N$, see equation
\eref{singlecorr}, the total power of $1/N$ is $\gamma$, the cyclicity of the
diagram.  Since our diagrams must be connected, the smallest cyclicity is
$\gamma=0$ if the diagram is a tree.  The trees can be generated recursively,
since by cutting a tree at the top node of degree $2l$ (after the root) we
obtain $2l-1$ subtrees, as illustrated in \fref{treecutting}.

To track the trees and their nodes, the generating function
$F(\bx,\bz_{i},\bz_{o})$ was introduced~\cite{bhn08} where the powers of 
\begin{itemize}
\item $x_l$ enumerate the number of $l$-encounters,
\item $z_{i, l}$ enumerate the number of $l$-encounters
  that $i$-touch the lead,
\item $z_{o, l}$ enumerate the number of $l$-encounters
  that $o$-touch the lead.
\end{itemize}
Later we will assign values to these variables which will produce the correct
semiclassical contributions of the trees.  Note that the contributions of the
links and leaves will be absorbed into the contributions of the nodes hence we
do not directly enumerate the links in the generating function $F$.  Inside $F$
we want to add all the possible trees and for each have a multiplicative
contribution of its nodes.  For example, the tree in
\fref{fourtrajectorystructures}i and its relatives in
\fref{fourtrajectorystructureleads} would contribute
\begin{equation}
 x_3 x_2 + z_{i,3} x_2 + x_3 z_{o,2} + z_{i,3} z_{o,2} =
\rbr{x_3+z_{i,3}}\rbr{x_2 + z_{o,2}} .
\end{equation}

A technical difficulty is that the top node may (if there are no further nodes)
be able to both $i$-touch and $o$-touch, but clearly not at the same time.  An
auxiliary generating function $f = f(\bx, \bz_{i}, \bz_{o})$ is thus introduced
with the restriction that the top node is not allowed to $i$-touch the lead.  We
denote by `empty' a tree which contains no encounter nodes (like
\fref{treecutting}d).  An empty tree is assigned the value 1 (\ie $f(0)=1$) to
not affect the multiplicative factors.  To obtain a recursion for $f$ we
separate the tree into its top node of degree $2l$ and $2l - 1$ subtrees as in
\fref{treecutting}.  As can be seen from the Figure, $l$ of the new trees (in
the odd positions from left to right) start with an incoming channel, while the
remaining $l-1$ even numbered subtrees start with an outgoing channel, and
correspond to a tree with the $i$'s and $o$'s are reversed.  For these we use
the generating function $\fh$ where the roles of the $\bz$ variables
corresponding to leaves of one type are switched so $\fh = f(\bx, \bz_{o},
\bz_{i})$.  The tree then has the contribution of the top node times that of all
the subtrees giving $x_l f^l \fh^{l-1}$.

The top node may also $o$-touch the lead, but for this to happen all the
odd-numbered subtrees must be empty (\ie they must contain no further nodes and
end directly in an outgoing channel).  When this happens we just get the
contribution of $z_{o,l}$ times that of the $l-1$ even subtrees:
$z_{o,l}\fh^{l-1}$.  In total we have
\begin{equation}
  \label{eq:recur_f}
  f = 1 + \sum_{l=2}^\infty \left[ x_l f^{l}
    {\fh}^{l-1} + z_{o,l} \fh^{l-1} \right] ,
\end{equation}
and similarly
\begin{equation}
  \label{eq:recur_fhat}
  \fh = 1 + \sum_{l=2}^\infty \left[ x_l {\fh}^{l}
    {f}^{l-1} + z_{i,l} f^{l-1} \right] .
\end{equation}
For $F$ we then reallow the top node to $i$-touch the lead which means that the
even subtrees must be empty and a contribution of $z_{i,l}f^{l}$, giving
\begin{equation}
 F =  f + \sum_{l=2}^\infty z_{i,l} f^{l} = \sum_{l=1}^\infty z_{i,l} f^{l} ,
  \label{eq:recur_F}
\end{equation}
if we let $z_{i,1} = 1$ (and also $z_{o,1} = 1$ for symmetry).  Picking an
$o$-leaf as the root instead of an $i$-leaf should lead to the same trees and
contributions so $F$ should be symmetric upon swapping $\bz_{i}$ with $\bz_{o}$
and $f$ with $\fh$.  These recursions enumerate all possible trees (which
represent all diagrams at leading order in inverse channel number) and we now
turn to evaluating their contributions to the correlation functions $C(\en,n)$.

\section{\label{singlelead}Density of states with a single lead}

To calculate the contribution of each diagram,
\ocites{rs02,heusleretal06,mulleretal07} used the ergodicity of the classical
motion to estimate how often the electron trajectories are likely to approach
each other and have encounters.  Combined with the sum rule~\cite{ha84,rs02} to
deal with the stability amplitudes, \ocite{heusleretal06} showed that the
semiclassical contribution can be written as a product of integrals over the
durations of the links and the stable and unstable separations of the stretches
in each encounter.  One ingredient is the survival probability that the electron
trajectories remain inside the system (these are followed by the holes whose
conditional survival probability is then 1) which classically decays
exponentially with their length and the decay rate $1/\tD=N/T_{\mathrm{H}}$.   A
small but important effect is that the small size of the encounters means the
trajectories are close enough to remain inside the system or escape (hit the
lead) together so only one traversal of each encounter needs to be counted in
the total survival probability
\begin{equation}
\exp\rbr{-\frac{N}{T_{\mathrm{H}}}t_{\mathrm{x}}},\qquad t_{\mathrm{x}} =
\sum_{i=1}^{L+n} t_{i} +\sum_{\alpha=1}^{V}t_{\alpha} ,
\end{equation}
where the $t_{i}$ are the durations of the ($n+L$) link stretches and
$t_{\alpha}$ the durations of the $V$ encounters so that the exposure time
$t_{\mathrm{x}}$ is shorter than the total trajectory time (which includes $l$
copies of each $l$-encounter).

As reviewed in \ocite{mulleretal07} the integrals over the links and the
encounters (with their action differences) lead to simple diagrammatic rules
whereby
\begin{itemize}
 \item each link provides a factor of $T_{\mathrm{H}}/\cbr{N\rbr{1-\rmi\en}}$ ,
 \item each $l$-encounter inside the cavity provides a factor of $-N\rbr{1-\rmi
l\en}/T_{\mathrm{H}}^{l}$ ,
\end{itemize}
with the $\rbr{1-\rmi l\en}$ deriving from the difference between the exposure
time and the total trajectory time.  Recalling the prefactor in
\eref{singlesemiclcorr} and that $L$ is the total number of links in the
encounters, it is clear that all the Heisenberg times cancel.  The channel
number factor $N^{-2n}$ from these rules and the prefactor (with $n=L-V+1$)
cancels with the sum over the channels in \eref{singlesemiclcorr} as each of the
$2n$ channels can be chosen from the $N$ possible channels (to leading order).  

With this simplification, each link gives $\rbr{1-\rmi\en}^{-1}$, each encounter
$-\rbr{1-\rmi l\en}$ and each leaf a factor of $1$.  To absorb the link
contributions into those of the encounters (nodes) we recall that the number of
links is $n+ \sum_{\alpha=1}^{V} l_\alpha$, where $\alpha$ labels the $V$
different encounters.  Therefore the total contribution factorizes as
\begin{equation}
\frac{1}{\rbr{1-\rmi\en}^n}\prod_{\alpha=1}^{V}\frac{-\rbr{1-\rmi
l_\alpha\en}}{\rbr{1-\rmi\en}^{l_\alpha}} .
 \label{singlesemicldiffchan}
\end{equation}
Moving an $l$-encounter into the lead, as in \fref{fourtrajectorystructureleads}
means losing that encounter, $l$ links and combining $l$ channels so we just
remove that encounter from the product above (or give it a factor 1 instead).

\subsection{Generating function}

Putting these diagrammatic rules into the recursions in \sref{treerecursions}
then simply means setting
\begin{equation}
x_{l} = \frac{-\rbr{1-\rmi l\en}}{\rbr{1-\rmi\en}^{l}} \cdot
\tilde{r}^{l-1},\qquad z_{i,l}=z_{o,l}=1\cdot \tilde{r}^{l-1} ,
\label{singlerecvalues}
\end{equation}
where we additionally include powers of $\tilde{r}$ to track the order of the
trees and later generate the semiclassical correlation functions.  The total
power of $\tilde{r}$ of any tree is $\sum_{l>1}(l-1)v_{l} = L-V = n-1$.  To get
the required prefactor of $\rbr{1-\rmi\en}^{-n}$ in \eref{singlesemicldiffchan}
we can then make the change of variable
\begin{equation}
 f=g(1-\rmi\en),\qquad\tilde{r}=\frac{r}{1-\rmi\en} ,
 \label{singlechangeofvar}
\end{equation}
so that the recursion relation \eref{eq:recur_f} becomes
\begin{equation}
  \label{eq:recur_g}
 g(1-\rmi\en) = 1 - \sum_{l=2}^\infty r^{l-1}g^{l}\gh^{l-1}(1-\rmi l\en) +
\sum_{l=2}^\infty r^{l-1}\gh^{l-1} ,
\end{equation}
and similarly for $\gh$.  Using geometric sums (the first two terms are the
$l=1$ terms of the sums) this is
\begin{equation}
\frac{g}{1-rg\gh} = \frac{\rmi\en g}{\rbr{1-rg\gh}^2} + \frac{1}{1-r\gh} .
\end{equation}
We note that the since $\fh$ is obtained from $f$ by swapping $\bz_i$ and
$\bz_o$ and in our substitution \eref{singlerecvalues} $\bz_i=\bz_o$, the
functions $\fh$ and $f$ are equal.  Taking the numerator of the equation above
and substituting $\gh=g$ leads to 
\begin{equation} \label{singlegeneq}
g-\frac{1}{1-\rmi\en}=\frac{rg^2}{1-\rmi\en}\cbr{g-1-\rmi\en} .
\end{equation}

To obtain the desired generating function of the semiclassical correlation
functions we set $F=G\rbr{1-\rmi\en}$ in \eref{eq:recur_F}, along with the other
substitutions in \eref{singlerecvalues} and \eref{singlechangeofvar},
\begin{equation}
 G(\en,r)=\frac{g}{1-rg}, \qquad G(\en,r)=\sum_{n=1}^{\infty} r^{n-1}C(\en,n) ,
 \label{singleGen}
\end{equation}
so that by expanding $g$ and hence $G$ in powers of $r$ we obtain all the
correlation functions $C(\en,n)$.  This can be simplified by rearranging
\eref{singleGen} and substituting into \eref{singlegeneq} to get the cubic for
$G$ directly
\begin{equation}
 r(r-1)^2G^3+r(3r+\rmi\en-3)G^2+(3r+\rmi\en-1)G + 1 = 0 .
 \label{singleGeneq}
\end{equation}

\subsection{\label{singledos}Density of states}

The density of states of a chaotic Andreev billiard with one superconducting
lead \eref{dossemieqn} can be rewritten as 
\begin{equation}
 d(\en)=1-2\mathrm{Im}\pdiff{}{\en}\sum_{n=1}^{\infty}\frac{(-1)^{n-1}C(\en,n)}{
n} ,
 \label{dossingle}
\end{equation}
where without the $1/n$ the sum would just be $G(\en,-1)$ in view of
\eref{singleGen}.  To obtain the $1/n$ we can formally integrate to obtain a new
generating function $H(\en,r)$,
\begin{eqnarray}
H(\en,r)&=&\frac{1}{\rmi r}\pdiff{}{\en}\int G(\en,r)\rmd r, \nonumber \\
H(\en,r)&=&\sum_{n=1}^{\infty} \frac{r^{n-1}}{\rmi n}\pdiff{C(\en,n)}{\en} ,
 \label{singleintgen}
\end{eqnarray}
so the density of states is given simply by
\begin{equation}
d(\en)=1-2\mathrm{Re}H(\en,-1) .
 \label{singleintgeneqn2}
\end{equation}
To evaluate the sum in \eref{dossingle} we now need to integrate the solutions
of \eref{singleGeneq} with respect to $r$ and differentiate with respect to
$\en$.  Since $G$ is an algebraic generating function, \ie the solution of an
algebraic equation, the derivative of $G$ with respect to $\en$ is also an
algebraic generating function~\cite{stanley01}.  However, this is not generally
true for integration, which can be seen from a simple example of $f=1/x$, which
is a root of an algebraic equation, unlike the integral of $f$.  Solving
equation \eref{singleGeneq} explicitly and integrating the result is also
technically challenging, due to the complicated structure of the solutions of
the cubic equations.  Even if it were possible, this approach would fail in the
presence of magnetic field, when $G$ is a solution of a quintic equation, see
\sref{singlemag}, or in the presence of a phase difference between two
superconductors.

The approach we took is to conjecture that $H(\en,r)$ is given by an algebraic
equation, perform a computer-aided search over equations with polynomial
coefficients and then prove the answer by differentiating appropriately.
We found that
\begin{eqnarray}
&& (\en r)^2(1-r)H^3+\rmi\en r[r(\rmi\en-2)+2(1-\rmi\en)]H^2 \nonumber \\
&& {} +[r(1-2\rmi\en)-(1-\rmi\en)^2]H+1=0 ,
 \label{singleHeq}
\end{eqnarray}
when expanded in powers of $r$, agrees for a range of values of $n$ with the
expansion of \eref{singleintgen} derived from the correlation functions obtained
from \eref{singleGeneq}.  To show that \eref{singleHeq} agrees with
\eref{singleintgen} to all orders in $r$ we use a differentiation algorithm to
find an equation for the intermediate generating function
\begin{eqnarray}
I(\en,r)&=&\frac{1}{\rmi}\pdiff{G(\en,r)}{\en}=\pdiff{[rH(\en,r)]}{r}, \nonumber
\\
I(\en,r)&=&\sum_{n=1}^{\infty} \frac{r^{n-1}}{\rmi}\pdiff{C(\en,n)}{\en} ,
 \label{singleintermediate}
\end{eqnarray}
both starting from (\ref{singleGeneq}) and from (\ref{singleHeq}) and verifying
that the two answers agree.

The differentiation algorithm starts with the algebraic equation for a formal
power series $\eta$ in the variable $x$ which satisfies an equation of the form
\begin{equation}
 \Phi(x,\eta)\mathrel{\mathop:}=p_0(x)+p_1(x)\eta+\ldots+p_m(x)\eta^m=0 ,
 \label{algebraic}
\end{equation}
where $p_0(x),\ldots,p_m(x)$ are some polynomials, not all of them zero.  The
aim is to find an equation satisfied by $\xi=\rmd\eta/\rmd x$, of the form
\begin{equation}
 q_0(x)+q_1(x)\xi+\ldots+q_m(x)\xi^m=0 ,
 \label{algebraic_xi}
\end{equation}
where $q_0(x),\ldots,q_m(x)$ are polynomials.  Differentiating \eref{algebraic}
implicitly yields
\begin{equation}
  \xi=-\frac{\partial \Phi(x,\eta)}{\partial x} \rbr{\frac{\partial
\Phi(x,\eta)}{\partial\eta}}^{-1}=\frac{P(\eta,x)}{Q(\eta,x)} ,
 \label{implicitdiff}
\end{equation}
where $P$ and $Q$ are again polynomial. After substituting this expression into
the algebraic equation for $\xi$ and bringing everything to the common
denominator we get
\begin{eqnarray}
q_0(x)Q^m(x,\eta)+q_1(x)P(x,\eta)Q^{m-1}(x,\eta)&& \nonumber \\
{} +\ldots +q_m(x)P^m(x,\eta)=0 . &&
 \label{fullinteq}
\end{eqnarray}
However, this equation should only be satisfied modulo the polynomial
$\Phi(x,\eta)$.  Namely, we use polynomial division and substitute
$P^j(x,\eta)Q^{m-j}(x,\eta)=T(x,\eta)\Phi(x,\eta)+R_j(x,\eta)$ into
\eref{fullinteq}.  Using 
\eref{algebraic} we arrive at
\begin{equation}
q_0(x)R_0(x,\eta)+q_1(x)R_1(x,\eta)+\ldots+q_m(x)R_m(x,\eta)=0 .
\label{reducedinteq}
\end{equation}
The polynomials $R_j$ are of degree of $m-1$ in $\eta$. Treating
\eref{reducedinteq} as an identity with respect to $\eta$ we thus obtain $m$
linear equations on the coefficients $q_j$. Solving those we obtain $q_j$ as
rational functions of $x$ and multiplying them by their common denominator gives
the algebraic equation for $\xi$.
\begin{figure*}
\subfigure{\includegraphics[width=0.48\textwidth]{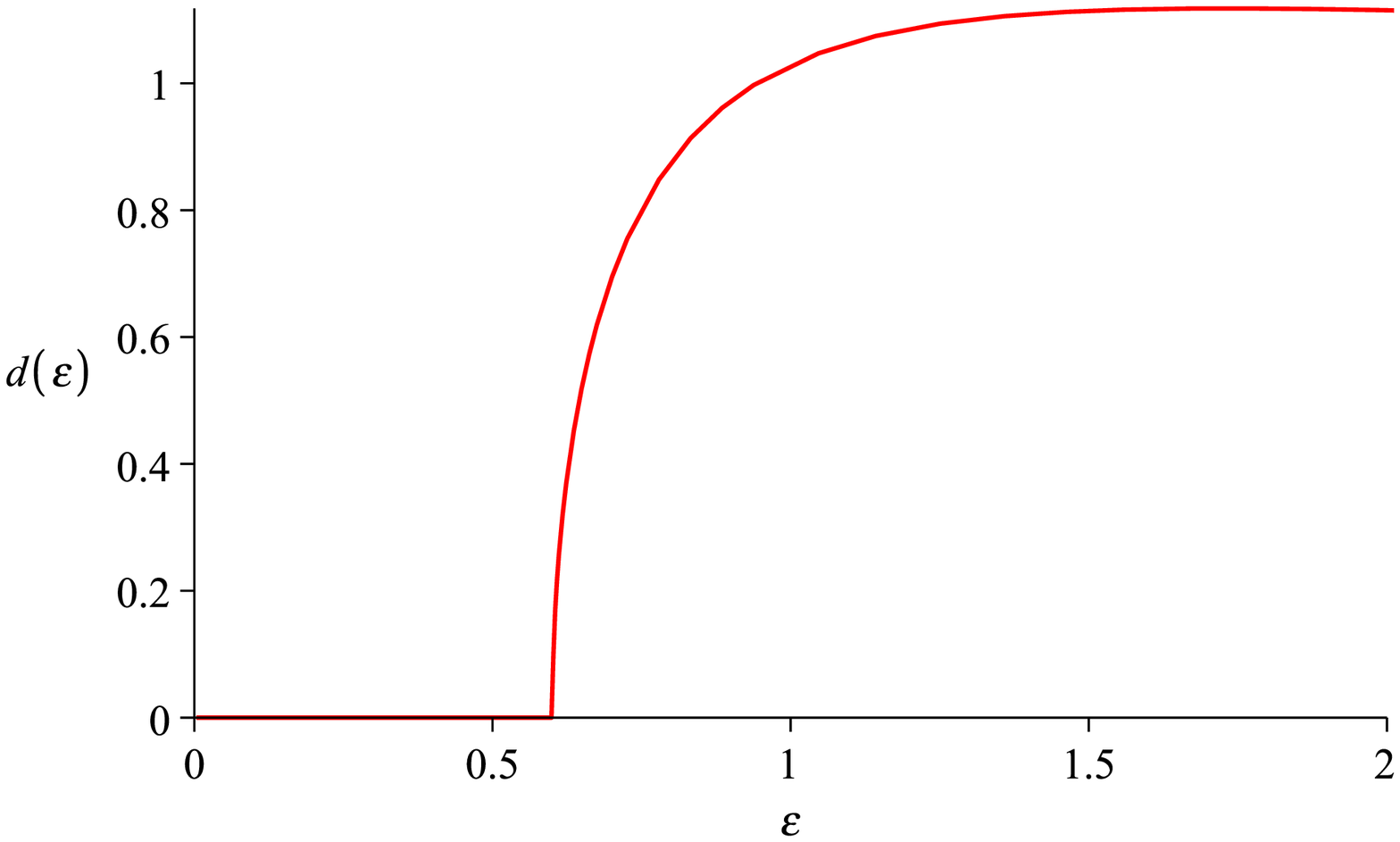}}
\subfigure{\includegraphics[width=0.48\textwidth]{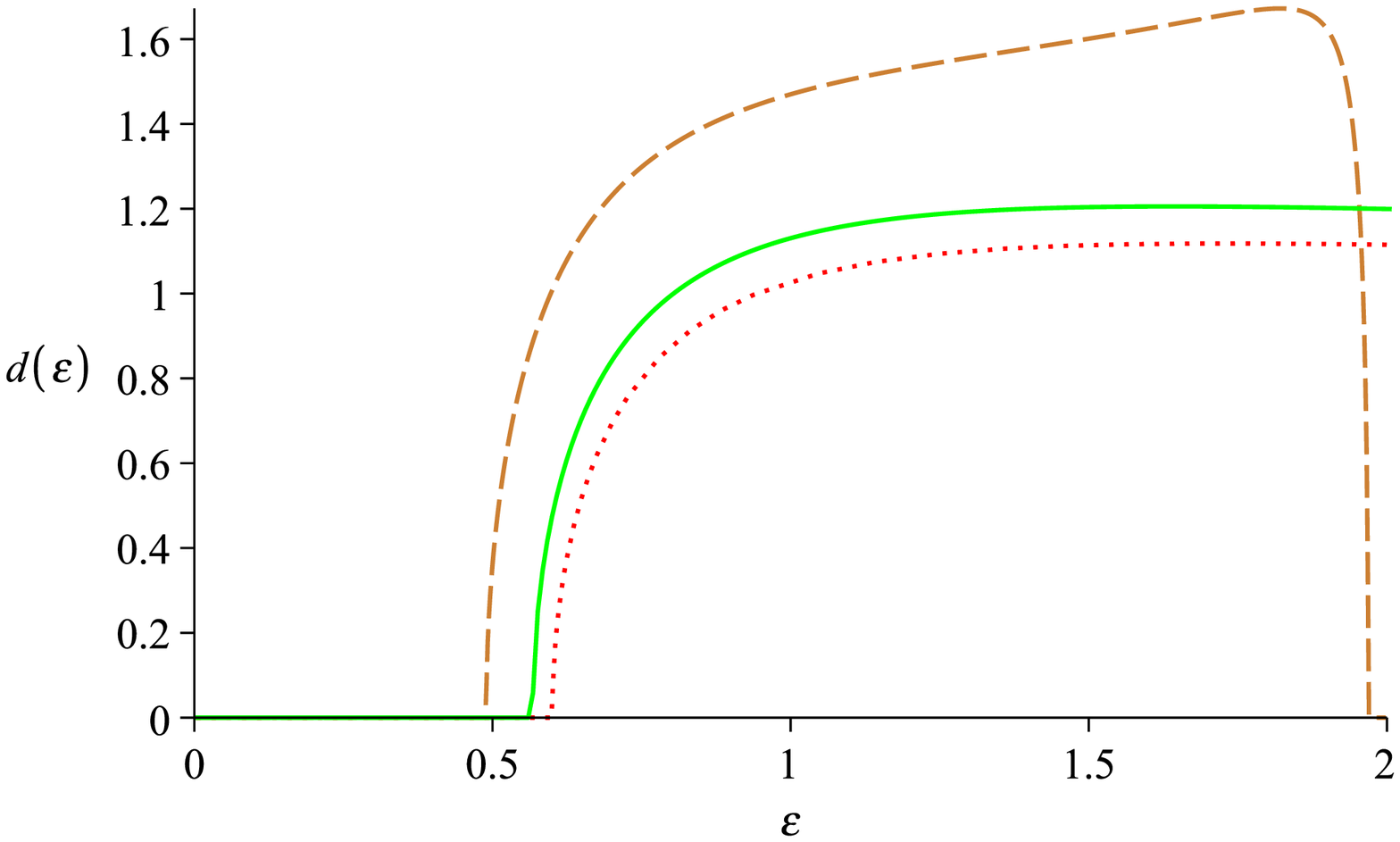}}
 \put(-486,0){(a)}
 \put(-238,0){(b)}
 \caption{\label{singledosplot}(a) The density of states of a chaotic quantum
dot coupled to a single superconductor at $E\ll\Delta$. (b) The density of
states with a finite bulk superconducting gap $\Delta=2E_\mathrm{T}$ (dashed
line) and $\Delta=8E_\mathrm{T}$ (solid line) compared to the previous case in
(a) with $\Delta\rightarrow\infty$ (dotted line).}
\end{figure*}

Performing this algorithm on $G$ from \eref{singleGeneq}, with $x=\rmi\en$, and
on $rH$ from \eref{singleHeq}, with $x=r$, leads to the same equation, given as
\eref{simpleIeqn} in the appendix, for the intermediate function defined in
\eref{singleintermediate} and therefore proves the validity of the equation
\eref{singleHeq}.  Setting $\en=0$ in \eref{singleHeq} then shows that
$\pdiff{C(\en,n)}{\en}\vert_{\en=0}=\rmi n$ as mentioned in \sref{semiapproach}.
To compare the final result \eref{singleintgeneqn2} with the RMT prediction we
can substitute $H(\en,-1)=\cbr{-\rmi W(\en)+1}/2$ into \eref{singleHeq}.  The
density of states is then given in terms of $W$ as $d(\en)=-\mathrm{Im}W(\en)$.
The equation for $W$ simplifies to the RMT result \eref{rmtsimplest}, and the
density of states then reads~\cite{melsenetal96}
\begin{equation}
 d(\en)=\left\{
 \begin{array}{ll}
  0 & \qquad \en\leq2\rbr{\frac{\sqrt{5}-1}{2}}^{5/2} \\
  \frac{\sqrt{3}}{6\en}\cbr{Q_+(\en)-Q_-(\en)} & \qquad
\en>2\rbr{\frac{\sqrt{5}-1}{2}}^{5/2}
 \end{array}\right. ,
 \label{singledossol}
\end{equation}
where $Q_{\pm}(\en)=\rbr{8-36\en^2\pm3\en\sqrt{3\en^4+132\en^2-48}}^{1/3}$. This
result is plotted in \fref{singledosplot}a and shows the hard gap extending up
to around $0.6E_{\mathrm{T}}$.

\subsection{Small bulk superconducting gap}

The calculation of the density of states above used the approximation that the
energy was well below the bulk superconductor gap, $E\ll\Delta$ or
$\en\ll\delta$ (for $\delta=\Delta/E_{\mathrm{T}}$), so that the phase shift at
each Andreev reflection was $\arccos(\en/\delta)\approx\pi/2$.  For higher
energies or smaller superconducting gaps, however, the density of states should
be modified~\cite{bb97} to
\begin{equation}
 d(\en)=1+\mathrm{Re}\frac{2}{\sqrt{\delta^2-\en^2}}+2\mathrm{Im}\sum_{n=1}^{
\infty}\pdiff{}{\en}\cbr{\frac{\alpha(\en)^{2n}C(\en,n)}{n}} ,
 \label{dossinglehighen}
\end{equation}
where $\alpha(\en)=\delta/(\en+\rmi\sqrt{\delta^2-\en^2})$ as in
\eref{alphadefeqn}. When taking the energy derivative in the sum in
\eref{dossinglehighen} we can split the result into two sums and hence two
contributions to the density of states
\begin{eqnarray}
 \label{dossinglehigheninter}
d(\en)&=&1+2\mathrm{Im}\sum_{n=1}^{\infty}\frac{\alpha(\en)^{2n}}{n}\pdiff{C(\en
,n)}{\en}  \\
&& {}
+\mathrm{Re}\frac{2}{\sqrt{\delta^2-\en^2}}\cbr{1+2\sum_{n=1}^{\infty}\frac{
\alpha(\en)^{2n}C(\en,n)}{n}} . \nonumber
\end{eqnarray}
Here the first term, which comes from applying the energy derivative to
$C(\en,n)$, gives an analogous contribution to the case $E\ll\Delta$ but with
$r=\alpha^2$ instead of $-1$ and involving $H(\en,\alpha^2)$ from
\eref{singleintgen} and \eref{singleHeq}. The second term in
\eref{dossinglehigheninter} comes from the energy derivative of $\alpha^{2n}$
and can be written using $G(\en,\alpha^2)$ from \eref{singleGen} and
\eref{singleGeneq}:
\begin{eqnarray}
 d(\en)&=&\mathrm{Re}\cbr{1+2\alpha^2 H(\en,\alpha^2)} \nonumber \\
 && {} +\mathrm{Re}\frac{2}{\sqrt{\delta^2-\en^2}}\cbr{1+2\alpha^2
G(\en,\alpha^2)} .
 \label{dossinglehighen2}
\end{eqnarray}
The effect of a finite bulk superconducting gap on the hard gap in the density
of states of the Andreev billiard is fairly small, for example as shown in
\fref{singledosplot}b even for $\delta=\Delta/E_{\mathrm{T}}=2$ the width just
shrinks to around $0.5E_{\mathrm{T}}$.  For $\delta=2$ the shape of the density
of states is changed somewhat (less so for $\delta=8$) and we can see just
before $\en=2$ it vanishes again giving a second thin gap.  This gap, and even
the way we can separate the density of states into the two terms in
\eref{dossinglehighen2}, foreshadows the effects of the Ehrenfest time (in
\sref{ehrenfest}).  For energies above the bulk superconducting gap
($\en>\delta$) we see a thin singular peak from the $\sqrt{\delta^2-\en^{2}}$
which quickly tends to the density of states of an Andreev billiard with an
infinite superconducting gap as the energy becomes larger.

\subsection{\label{singlemag}Magnetic field}

If a magnetic field is present, the time reversal symmetry is broken and we wish
to treat this transition semiclassically as in \ocites{tr03,sn06}.  Note that
since for the leading order diagrams each stretch is traversed in opposite
directions by an electron and a hole we are effectively considering the same
situation as for parametric correlations~\cite{nagaoetal07,ks07a}.   Either way,
the idea behind the treatment is that the classically small magnetic field
affects the classical trajectories very little, but adds many essentially random
small perturbations to the action.  The sum of these fluctuations is
approximated using the central limit theorem, and leads to an exponential
damping so the links now provide a factor of $T_{\mathrm{H}}/N(1-\rmi\en+b)$. 
The parameter $b$ is related to the magnetic field via
$b=\rbr{\Phi/\Phi_{\mathrm{c}}}^2$ as in \sref{rmt}. For an $l$-encounter
however, as the stretches are correlated and affected by the magnetic field in
the same way, the variance of the random fluctuations of all the stretches is
$l^2$ that of a single stretch. Hence each encounter now contributes
$N\rbr{1-\rmi l\en+l^2b}/T_{\mathrm{H}}^{l}$ and again the correlation inside
the encounters leads to a small but important effect.
\begin{figure}
 \includegraphics[width=\columnwidth]{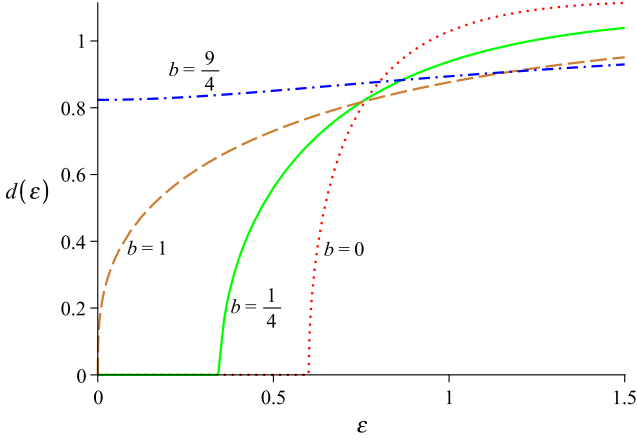}
 \caption{\label{singledosmagplot}The effect of a time reversal symmetry
breaking magnetic field on the density of states of a chaotic Andreev billiard
with a single superconducting lead for $b=0$ (dotted line), $b=1/4$ (solid
line), $b=1$ (dashed line) and $b=9/4$ (dashed dotted line).}
\end{figure}

Similarly to the treatment without the magnetic field above, we can put these
contributions into the recursions in \sref{treerecursions} by setting
\begin{equation}
x_{l} = \frac{-\rbr{1-\rmi l\en+l^2 b}}{\rbr{1-\rmi\en+b}^{l}} \cdot
\tilde{r}^{l-1},\qquad z_{i,l}=z_{o,l}=1\cdot \tilde{r}^{l-1} ,
\end{equation}
and
\begin{equation}
 f=g(1-\rmi\en+b),\qquad\tilde{r}=\frac{r}{1-\rmi\en+b} .
 \label{singlemagchangeofvar}
\end{equation}
The intermediate generating function is then given by the implicit equation
\begin{eqnarray}
&& -r^2g^5+(1+\rmi\en+b)r^2g^4+(2-\rmi\en-b)rg^3 \nonumber \\
&& {} -(2+\rmi\en-b)rg^2-(1-\rmi\en+b)g+1=0 ,
  \label{singlemaggeq}
\end{eqnarray}
and the generating function $G(\en,b,r)$ of the magnetic field dependent
correlation functions $C(\en,b,n)$, which is still connected to $g$ via
$G=g/(1-rg)$, is given by
\begin{eqnarray}
&& r^2(r-1)^3G^5\nonumber \\
&& {} +\rbr{\rmi\en r-\rmi\en+5r^2-10r+5-br-b}r^{2}G^4\nonumber\\
&& {} +\rbr{3\rmi\en r-\rmi\en+10r^2-12r+2-3br-b}rG^3 \nonumber\\
&& {} +(3\rmi\en+10r-6-3b)rG^2\nonumber \\
&& {} -(1-5r-\rmi\en+b)G+1=0 . 
 \label{singlemagGeq}
\end{eqnarray}
Removing the magnetic field by setting $b=0$ reduces both of these equations (after
factorizing) to the previous results~\eref{singlegeneq} and~\eref{singleGeneq}. 
Next we again search for and verify an algebraic equation for
$H(\en,b,r)=1/(\rmi r)\int[\partial G(\en,b,r)/\partial\en]\rmd r$, though the
higher order makes this slightly more complicated, finding
\begin{eqnarray}
&& 4b^2r^4\rbr{r-1}H^5+4br^3\cbr{\rmi\en-3b+r\rbr{2b-\rmi\en}}H^4 \nonumber\\
&& {} +r^2\left[\en^2\rbr{1-r}+2\rmi\en b\rbr{5-3r}-b\rbr{13b+4}\right.
\nonumber \\
&& \qquad {} +br\rbr{5b+4}\big]H^3 \nonumber \\
&& {} +r\Big[2\rbr{\rmi\en-3b}\rbr{1-\rmi\en+b} \label{singlemagHeq}\\
&& \qquad \left. {} +r\rbr{\rbr{1-\rmi\en+b}^2+4b-1}\right]H^2 \nonumber\\
&& {} -\cbr{\rbr{1-\rmi\en+b}^2-r\rbr{1-2\rmi\en+2b}}H+1=0 . \nonumber
\end{eqnarray}
In order to check the agreement with the RMT result we substitute
$H(\en,b,-1)=\cbr{-\rmi W(\en,b)+1}/2$ into \eref{singlemagHeq}. This leads to
\begin{eqnarray}
&& b^2W^5-2b\en W^4-\rbr{4b-b^2-\en^2}W^3+2(2-b)\en W^2 \nonumber\\
&& {}+\rbr{4-4b+\en^2}W+4\en=0 ,
 \label{singlemagWeq}
\end{eqnarray}
which corresponds to the RMT result~\eref{origrmteq} with no phase ($\phi=0$).
The density of states calculated from this equation is shown in
\fref{singledosmagplot} for different values of $b$. The gap reduces for
increasing $b$, closes exactly at the critical flux ($b=1$) and the density of
states becomes flat (at $1$) as $b\to\infty$.

\section{\label{twoleads}Density of states with two leads}

\begin{figure}
 \includegraphics[width=\columnwidth]{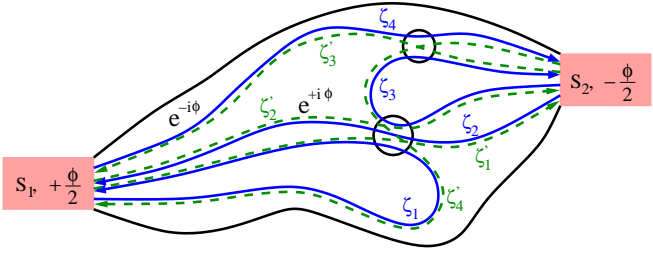}
 \caption{\label{twoleadpaths}The paths may start and end in either of the two
leads as shown. $\zeta_4$ as it travels from lead 1 to lead 2 obtains a phase
factor $\exp(-\rmi\phi)$, $\zeta_2$ traveling back contributes $\exp(\rmi\phi)$
while the others does not contribute any phase. The encounters are again marked
by circles and S$_1$ and S$_2$ denote the two superconducting leads at the
corresponding superconducting phases $\pm\phi/2$.  This diagram is equivalent to
the one in \fref{fourtrajectorystructures}f.}
\end{figure}

Next we consider a classically chaotic quantum dot connected to two
superconductors with $N_1$ and $N_2$ channels respectively and a phase
difference $\phi$, as depicted in \fref{andreevgeneralbilliard}a. The density of
states, as in \sref{rmt} and \ocites{melsenetal97,bb97}, can then be reduced to
equation \eref{dossemieqn} but with
\begin{equation}
 C(\en,\phi,n)=\frac{1}{N}\mathrm{Tr}\cbr{S^{*}\rbr{-\frac{\en\hbar}{2\tD}}\rme^
{-\rmi\tilde{\phi}}S\rbr{+\frac{\en\hbar}{2\tD}}\rme^{\rmi\tilde{\phi}}}^n ,
 \label{phasecorreqn}
\end{equation}
where $\tilde{\phi}$ is again a diagonal matrix whose first $N_{1}$ elements
from the first superconductor S$_1$ are $\phi/2$ and the remaining $N_{2}$
elements from S$_2$ are $-\phi/2$.  Note that the case $\phi=0$ corresponds to
the previous case of a single superconductor with $N=N_1+N_2$ channels.  When we
substitute the semiclassical approximation for the scattering matrix
\eref{semiclscat} into \eref{phasecorreqn}, and especially if we write the
scattering matrix in terms of its reflection and transmission subblocks, the
effect of the superconductors' phase difference becomes simple.  Namely, each
electron (unprimed) trajectory which starts in lead 1 and ends in lead 2 picks
up the phase factor  $\exp(-\rmi\phi)$ while each unprimed trajectory going from
lead 2 to lead 1 receives the factor $\exp(\rmi\phi)$.  Reflection trajectories
which start and end in the same lead have no additional phase factor, as
depicted in \fref{twoleadpaths}.  Since exchanging the leads gives the opposite
phase, we expect the solution to be symmetric if we simultaneously exchange
$N_1$ with $N_2$ and change $\phi$ to $-\phi$.

As these factors are multiplicative, we can equivalently say that each electron
trajectory leaving superconductor 1 or 2 picks up $\exp(-\rmi\phi/2)$ or
$\exp(\rmi\phi/2)$ while each one entering lead 1 or 2 picks up
$\exp(\rmi\phi/2)$ or $\exp(-\rmi\phi/2)$.  To include these factors in our
semiclassical diagrams, we can simply remember that in our tree recursions in
\sref{treerecursions} the channels we designated as `incoming' channels have
electrons leaving them while electrons always enter the outgoing channels.  Each
incoming channel (in the original channel sum in \eref{singlesemiclcorr}) can
still come from the $N$ possible channels, but with the trajectory leaving it now
provides the factor $N_{1}\exp(-\rmi\phi/2)+N_{2}\exp(\rmi\phi/2)$.  Similarly
each outgoing channel now provides the complex conjugate of this factor. 
Recalling the power of $N^{-2n}$ coming from the links and encounters, we can
update the contribution of each diagram or tree \eref{singlesemicldiffchan} to
\begin{eqnarray}
&&
\frac{\rbr{N_{1}\rme^{-\frac{\rmi\phi}{2}}+N_{2}\rme^{\frac{\rmi\phi}{2}}}^n\rbr
{N_{1}\rme^{\frac{\rmi\phi}{2}}+N_{2}\rme^{-\frac{\rmi\phi}{2}}}^n}{N^{2n}\rbr{
1-\rmi\en}^n}\nonumber\\
&& {} \times \prod_{\alpha=1}^{V}\frac{-\rbr{1-\rmi
l_\alpha\en}}{\rbr{1-\rmi\en}^{l_\alpha}} .
 \label{twoleadssemicl}
\end{eqnarray}
However, moving an $l$-encounter into lead 1 means combining $l$ incoming
channels, $l$ links and the encounter itself.  These combined incoming channels,
with $l$ electron trajectories leaving, will now only give the factor
$N_{1}\exp(-\rmi l\phi/2)+N_{2}\exp(\rmi l\phi/2)$ where the important
difference is that $l$ is inside the exponents.  We therefore make the
replacement
\begin{equation}
\frac{\rbr{N_{1}\rme^{-\frac{\rmi\phi}{2}}+N_{2}\rme^{\frac{\rmi\phi}{2}}}^l}{N^
{l}} \to \frac{\rbr{N_{1}\rme^{-\frac{\rmi l\phi}{2}}+N_{2}\rme^{\frac{\rmi
l\phi}{2}}}}{N}
 \label{twoleadsincomingreplacement}
\end{equation}
as well as removing the encounter from \eref{twoleadssemicl}.  Similarly when we
move the encounter into the outgoing leads we take the complex conjugate of
\eref{twoleadsincomingreplacement}.

To mimic these effects in the semiclassical recursions we can set
\begin{eqnarray}
&& x_{l} = \frac{-\rbr{1-\rmi l\en}}{\rbr{1-\rmi\en}^{l}} \cdot
\tilde{r}^{l-1},\nonumber \\
&&
\beta=\frac{\rbr{N_{1}\rme^{-\frac{\rmi\phi}{2}}+N_{2}\rme^{\frac{\rmi\phi}{2}}}
}{N} ,
 \label{phasechangeofvar1} \\
&& z_{i,l}=\frac{\rbr{N_{1}\rme^{-\frac{\rmi l\phi}{2}}+N_{2}\rme^{\frac{\rmi
l\phi}{2}}}}{N{\beta}^l} \cdot \tilde{r}^{l-1}, \nonumber \\
&& z_{o,l}=\frac{\rbr{N_{1}\rme^{\frac{\rmi l\phi}{2}}+N_{2}\rme^{-\frac{\rmi
l\phi}{2}}}}{N\rbr{\beta^{*}}^l} \cdot \tilde{r}^{l-1} ,
 \label{phasechangeofvar2} \\
&& f=g\frac{\rbr{1-\rmi\en}}{\beta \beta^{*}},\qquad\tilde{r}=r\frac{\beta
\beta^{*}}{\rbr{1-\rmi\en}} ,
 \label{phasechangeofvar3}
\end{eqnarray}
in \sref{treerecursions}.  Including these substitutions in the recursion
relation \eref{eq:recur_f} and summing we obtain
\begin{eqnarray}
\label{phasegeqn}
\frac{g}{\beta\beta^{*}-rg\gh} &=& \frac{\rmi\en\beta\beta^{*}
g}{\rbr{\beta\beta^{*}-rg\gh}^2} +
\frac{N_{1}}{N}\frac{1}{\beta^{*}\rme^{-\frac{\rmi\phi}{2}}-r\gh} \nonumber \\
&& {} + \frac{N_{2}}{N}\frac{1}{\beta^{*}\rme^{\frac{\rmi\phi}{2}}-r\gh} ,
\end{eqnarray}
and a similar equation from \eref{eq:recur_fhat}.  The generating function of
the correlation functions $C(\en,\phi,n)$ is then given from \eref{eq:recur_F}
by
\begin{equation}
\label{Gphaseeqn}
G =  \frac{N_{1}}{N}\frac{g}{\beta\rme^{\frac{\rmi\phi}{2}}-rg} +
\frac{N_{2}}{N}\frac{g}{\beta\rme^{\frac{-\rmi\phi}{2}}-rg} .
\end{equation} 

Returning to \eref{phasegeqn} and multiplying through by $\gh$, we can see that
the first two terms are symmetric in $g$ and $\gh$.  Combining the other two and
taking the difference from the corresponding equation for $\gh$ we have
\begin{eqnarray}
&&\frac{\gh\cbr{\rbr{\beta^{*}}^{2}
-r\gh}}{\rbr{\beta^{*}\rme^{-\frac{\rmi\phi}{2}}-r\gh}\rbr{\beta^{*}\rme^{\frac{
\rmi\phi}{2}}-r\gh}} \nonumber\\
 &=& \frac{g\cbr{\beta^{2}
-rg}}{\rbr{\beta\rme^{\frac{\rmi\phi}{2}}-rg}\rbr{\beta\rme^{-\frac{\rmi\phi}{2}
}-rg}} . \label{gghatphaseeqn}
\end{eqnarray}
The resulting quadratic equation, when substituted back into \eref{phasegeqn}
leads to a sixth order equation for $g$.  Note that the right hand side of
\eref{gghatphaseeqn} is (recalling \eref{phasechangeofvar1} and that
$N_1+N_2=N$) the same as \eref{Gphaseeqn} so it is clear that $G$ satisfies the
required symmetry upon swapping the leads (\ie swapping $N_1$ with $N_2$ and
$\phi$ with $-\phi$).

\subsection{Equal leads}

\begin{figure}
 \includegraphics[width=\columnwidth]{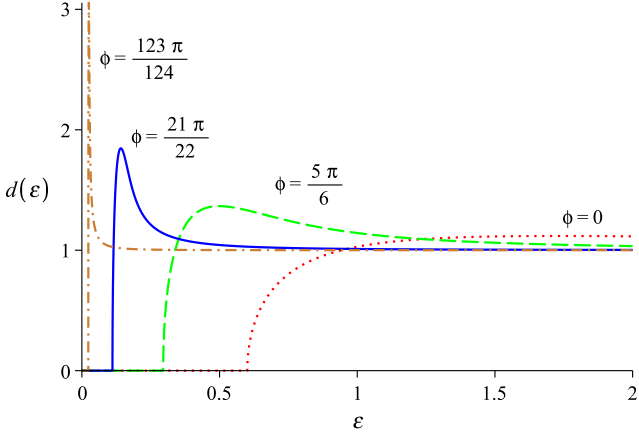}
 \caption{\label{dosplot}The density of states of a chaotic quantum dot coupled
to two superconductors with the same numbers of channels and phase differences
$0$ (dotted line), $5\pi/6$ (solid line), $21\pi/22$ (dashed line) and
$123\pi/124$ (dashed dotted line).}
\end{figure}

To make the equations more manageable we focus for now on the simpler case in which
the leads have equal size and $N_1=N_2=N/2$.  Then $\beta=\cos(\phi/2)$ is real
and we can see from \eref{gghatphaseeqn} or $\bz_{i}=\bz_{o}$ that $g=\gh$ is a
solution.  Putting this simplification into \eref{phasegeqn} we can obtain the
following quartic
\begin{equation}
 r^2g^4-r(1+r+\rmi\en r)g^3+2\rmi\en \beta^{2}
rg^2+(1-\rmi\en+r)\beta^{2}g-\beta^{4}=0 .
 \label{galg}
\end{equation}
We may also find an algebraic equation of fourth order for $G$ if we solve
\eref{Gphaseeqn} for $g$ and substitute the solution
\begin{equation}
 g=\frac{\beta}{2}\frac{2r\beta G+\beta -
\sqrt{\beta^{2}+4rG\rbr{1+rG}\rbr{\beta^2-1}}}{r(1+rG)} ,
 \label{gGrel}
\end{equation}
into \eref{galg}.  Note that we take the negative square root to agree with the
previous result when the phase is 0 (\ie $\beta=1$) though this sign does not
affect the equation one finally finds for $G$.  After the fourth order equation
for $G$ has been found we can again search for and verify an equation for
$H(\en,\phi,r)=1/(\rmi r)\int (\partial G(\en,\phi,r)/\partial\en)\rmd r$,
\begin{eqnarray}
&& \en^2r^3\cbr{1-2r\rbr{2\beta^2-1}+r^2}H^4 \nonumber\\
&& {} +\rmi\en r^2\left[2-3\rmi\en-4r\rbr{1-\rmi\en}\rbr{2\beta^2-1} \right.
\nonumber \\
&& \left. \qquad {} +r^2 \rbr{2-\rmi\en}\right]H^3\nonumber\\
&& {} - r\left[1-4\rmi\en-3\en^2-2r\rbr{1-3\rmi\en-\en^2}\rbr{2\beta^2-1}
\right. \nonumber \\
&&  \left. \qquad {} +r^2 \rbr{1-2\rmi\en}\right]H^2\nonumber\\
&& {} -\cbr{\rbr{1-\rmi\en}^2-2r\rbr{1-\rmi\en}\rbr{2\beta^2-1}+r^2}H \nonumber
\\
&& {} +\beta^2=0 . 
 \label{Hphaseeq}
\end{eqnarray}
In order to see the agreement of our result with the RMT prediction we again
substitute $H(\en,\phi,-1)=[-\rmi W(\en,\phi)+1]/2$ such that
$d(\en)=-\mathrm{Im}W(\en,\phi)$. If we do so we find
\begin{eqnarray}
&& \en^2\beta^2W^4+4\en \beta^2W^3+(4\beta^2-\en^2+2\en^2\beta^2)W^2 \nonumber
\\
&& {}+4\en\beta^2 W-\en^2+\en^2\beta^2=0 ,
 \label{semiphase}
\end{eqnarray}
which corresponds to \eref{origrmteq} for zero magnetic field. Moreover, if the
phase difference is zero (and $\beta=1$), we can take out the factor $W$ and
recover \eref{rmtsimplest}.

Solving this equation yields the density of states. If we insert different
values for the phase $\phi$ one finds that the hard gap in the density of states
decreases with increasing phase difference while the density of states has a
peak at the end of the gap which increases and becomes sharper with increasing
phase. Finally when the phase difference is equal to $\pi$ the gap closes and
the peak vanishes so the density of states becomes identical to $1$. This can
all be seen in \fref{dosplot}.

\subsection{\label{twomag}Magnetic field.}

\begin{figure}
 \includegraphics[width=\columnwidth]{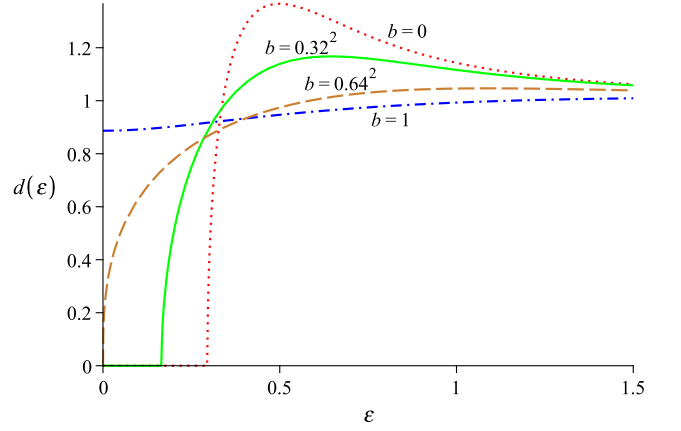}
\caption{\label{magphase1}Magnetic field dependence of the density of states of
a chaotic Andreev billiard with phase difference $\phi=5\pi/6$ for $b=0$ (dotted
line), $b=0.1024$ (solid line), $b=0.4096$ (dashed line) and $b=1$ (dashed
dotted line).}
\end{figure}
\begin{figure*}
 \subfigure{\includegraphics[width=0.28\textwidth]{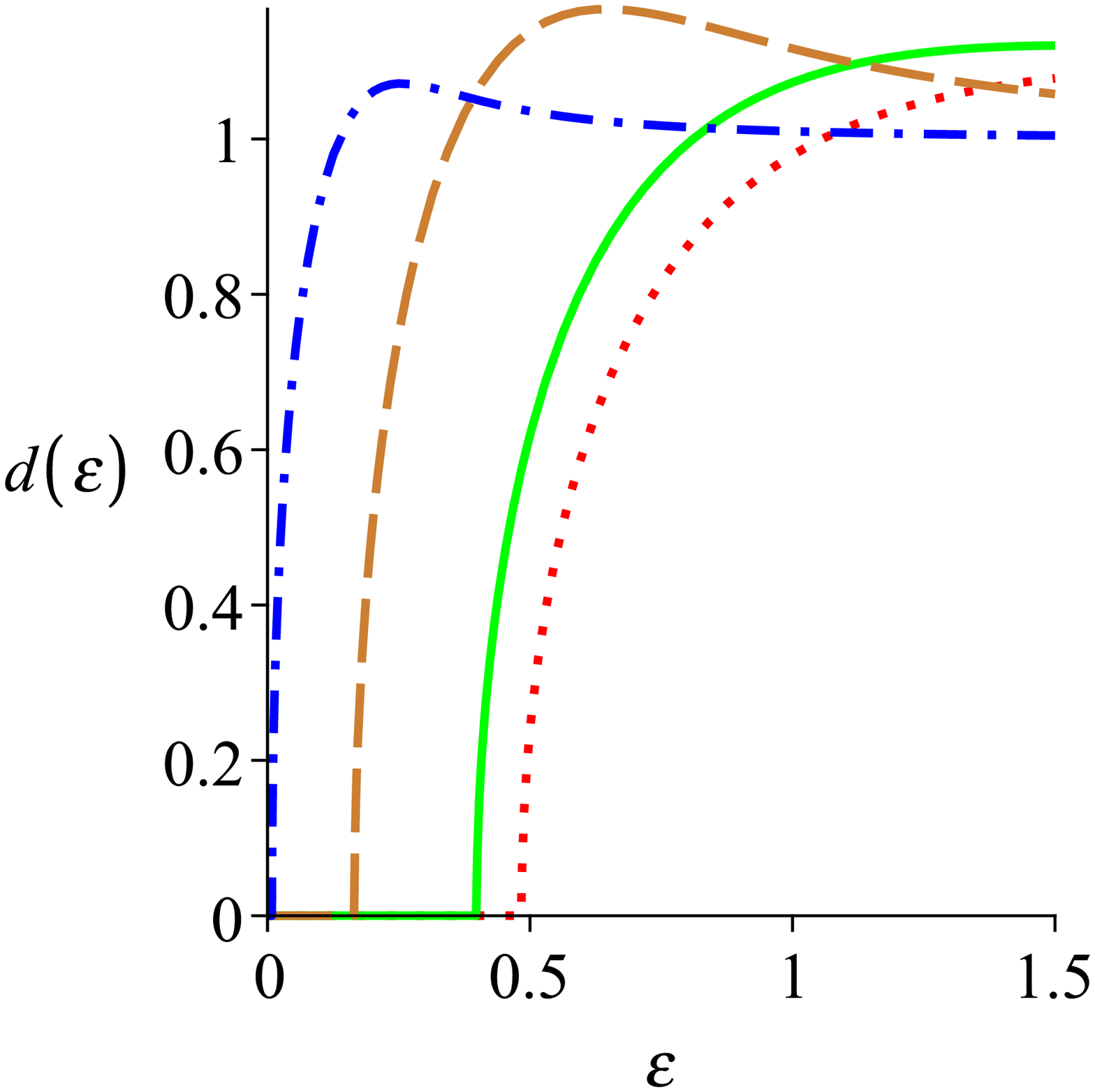}}
 \subfigure{\includegraphics[width=0.28\textwidth]{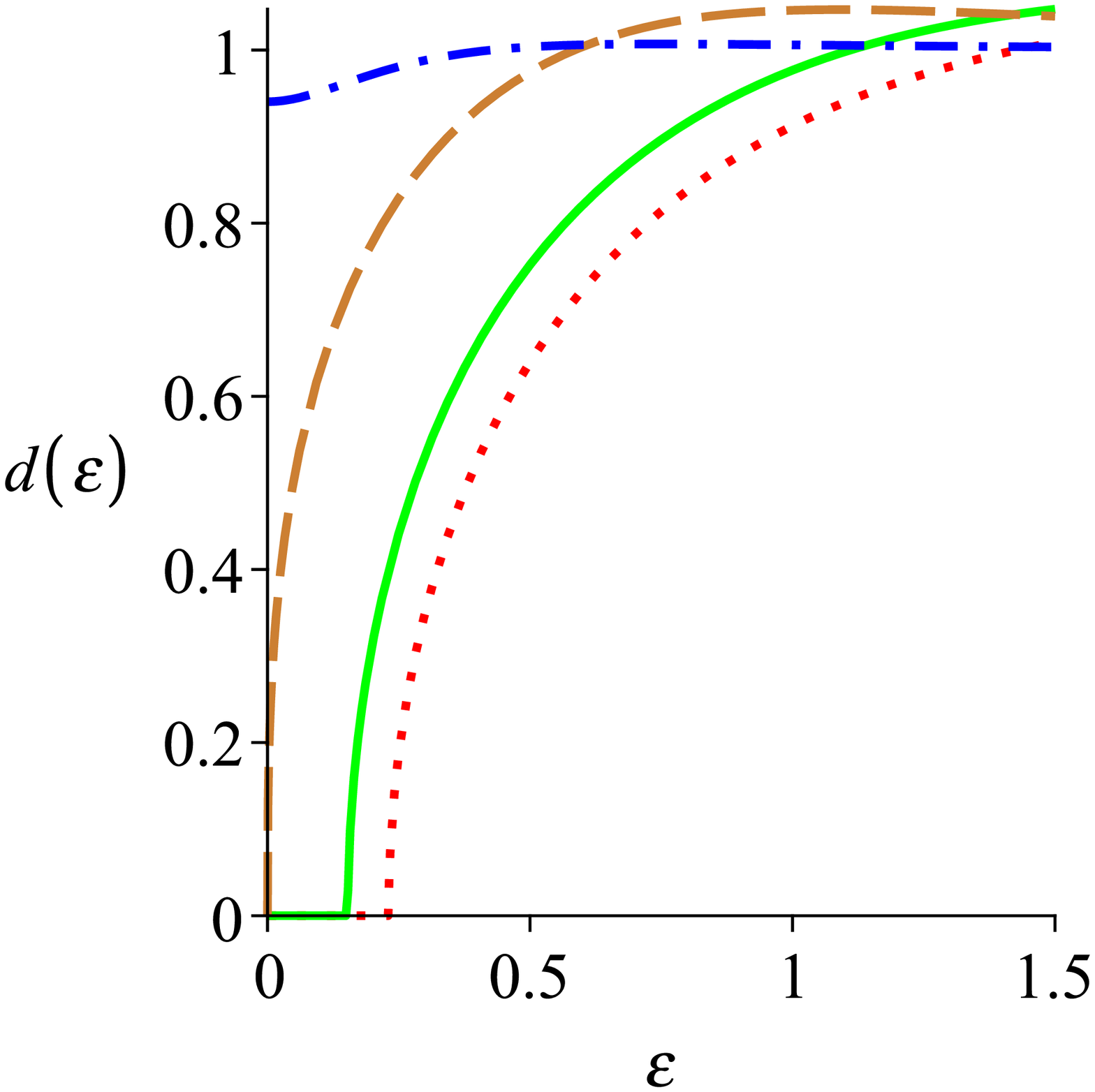}}
 \subfigure{\includegraphics[width=0.28\textwidth]{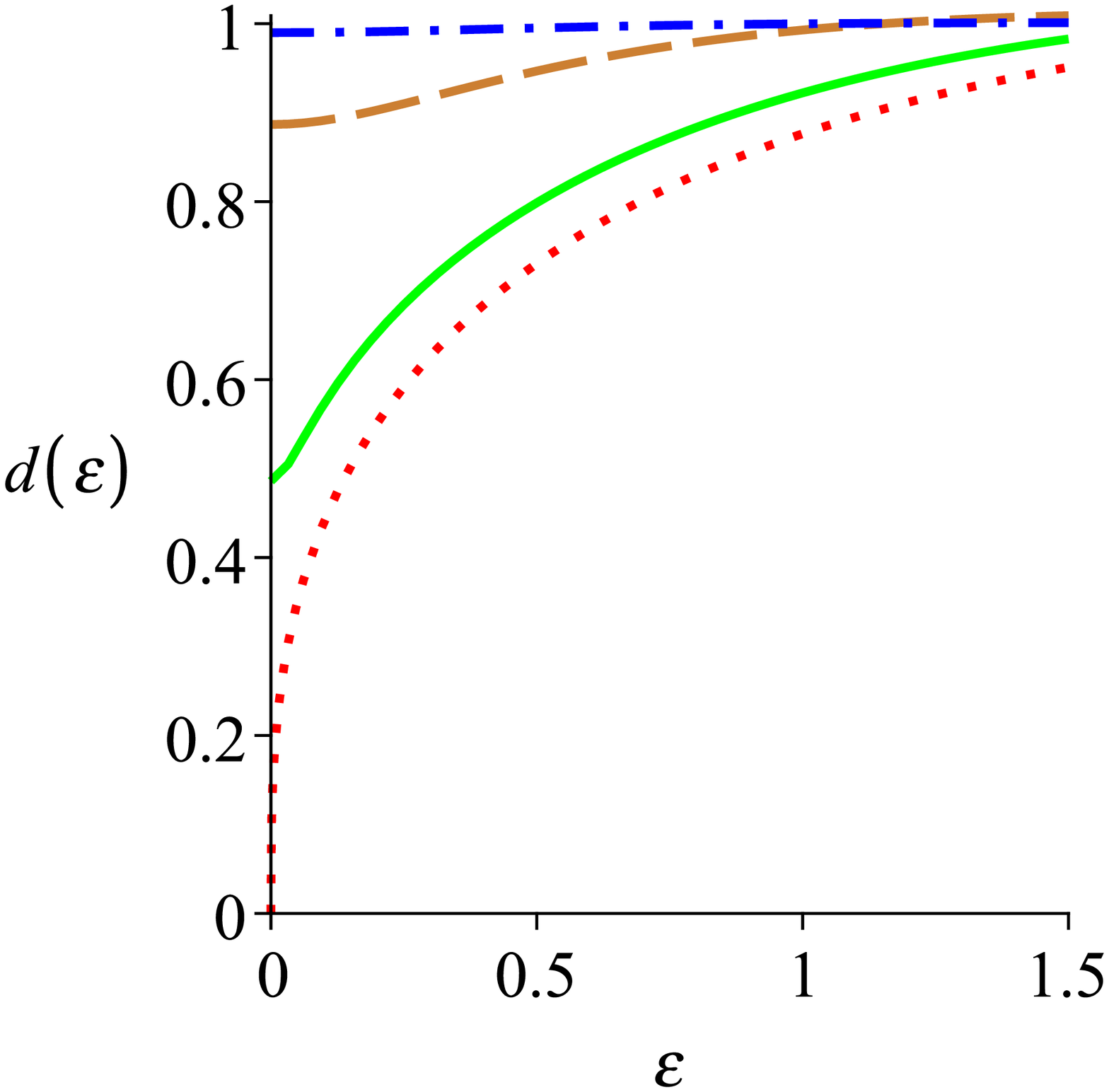}}
 \put(-430,0){(a)}
 \put(-282,0){(b)}
 \put(-134,0){(c)}
 \caption{\label{magphase2}Phase dependence of the density of states of a
chaotic Andreev billiard with phase difference $\phi=0$ (dotted line),
$\phi=\pi/2$ (solid line), $\phi=5\pi/6$ (dashed line) and $\phi=21\pi/22$
(dashed dotted line). (a) At magnetic field $b=0.1024$, (b) at $b=0.4096$ and
(c) at $b=1$.}
\end{figure*}

In the presence of a magnetic field, we again have to change the diagrammatic
rules as in \sref{singlemag}. Doing the calculation above with these modified
diagrammatic rules leads to a sixth order equation for $g$:
\begin{eqnarray}
&& r^3g^6-r^2\cbr{1+r\rbr{1+\rmi\en+b}}g^5-r^2\beta^2\rbr{1-2\rmi\en-2b}g^4
\nonumber\\
&& {}+r\beta^2\cbr{2-\rmi\en-b+r\rbr{2+\rmi\en-b}}g^3 \nonumber \\
&& {}-r\beta^4\rbr{1+2\rmi\en-2b}g^2 \nonumber\\
&& {}-\beta^4\rbr{1+r-\rmi\en+b}g+\beta^6=0 .
 \label{maggeq}
\end{eqnarray}
The relation \eref{Gphaseeqn} between $G$ and $g$ remains unchanged and
therefore we may find a sixth order equation for G.  We find the corresponding
$H$, which is recorded as \eref{magphaseHeqn} in the appendix, using a computer
search over sixth order equations with polynomial (in $\en$, $\phi$, $b$ and
$r$) coefficients whose expansion in $r$ \eref{singleintgen} matches the
correlation functions calculated by expanding $G$.  We note that for this order
polynomial it was not feasible (in terms of computational time and memory) to
solve the equations resulting from the differentiation algorithm described in
\sref{singledos} and to find the intermediate generating function $I$ in all
generality.  However, we succeeded in finding a polynomial equation for $I$ that
was satisfied by the derivatives of both $rH$ and $G$ for a large number of
numerical values of the parameters $(\en,\phi,b)$.  For each parameter involved,
the number of the values checked was larger than the maximum degree of the
parameter in the conjectured equation. While we cannot rule out the possibility
that the true equation for $I$ has a higher order, given the large number of
numerical values checked this is highly unlikely.

From $H$ we obtain the equation for $W(\en,\phi,b)$,
\begin{eqnarray}
&& b^2\beta^2W^6-2\en b\beta^2
W^5+\rbr{2b^2\beta^2+\en^2\beta^2-4b\beta^2-b^2}W^4 \nonumber\\
&& {} +2\rbr{\en b+2\en\beta^2-2\en b\beta^2}W^3 \nonumber \\
&& {}+\rbr{4\beta^2-b^2-\en^2-4b\beta^2+b^2\beta^2+2\en^2\beta^2}W^2 \nonumber\\
&& {}+2\rbr{\en b+2\en\beta^2-\en b\beta^2}W-\en^2+\en^2\beta^2=0 ,
 \label{magWeqn}
\end{eqnarray}
which corresponds exactly to the full RMT result \eref{origrmteq} expanded.

As an example, the magnetic field dependence of the density of states is shown
at the phase difference of $5\pi/6$ in \fref{magphase1}.  As the magnetic field
is increased one finds a reduction of the gap and the peak appearing for a phase
difference $\phi>0$ vanishes again. Moreover the higher the phase difference, the
lower the magnetic field needed to close the gap. While for $\phi=0$
the gap closes at $b=1$ in the case of a phase difference of $5\pi/6$ one needs
$b\approx0.4096$ and for $\phi=21\pi/22$ a magnetic field corresponding to
$b\approx0.1024$ closes the gap. In particular the critical magnetic field for
which the gap closes is given by~\cite{melsenetal97}
\begin{equation}
 b_c=\frac{2\cos\rbr{\phi/2}}{1+\cos\rbr{\phi/2}} .
 \label{criticalmagfield}
\end{equation}
For ever increasing magnetic field the density of states approaches 1 and we can
see that a higher phase difference causes a faster convergence to this limit. 
Some examples are plotted in \fref{magphase2} and there we see that for $b=1$
the curve for $\phi=21\pi/22$ is nearly constant.

\subsection{\label{unequal}Unequal leads}

\begin{figure*}
\centering
 \subfigure{\includegraphics[width=0.28\textwidth]{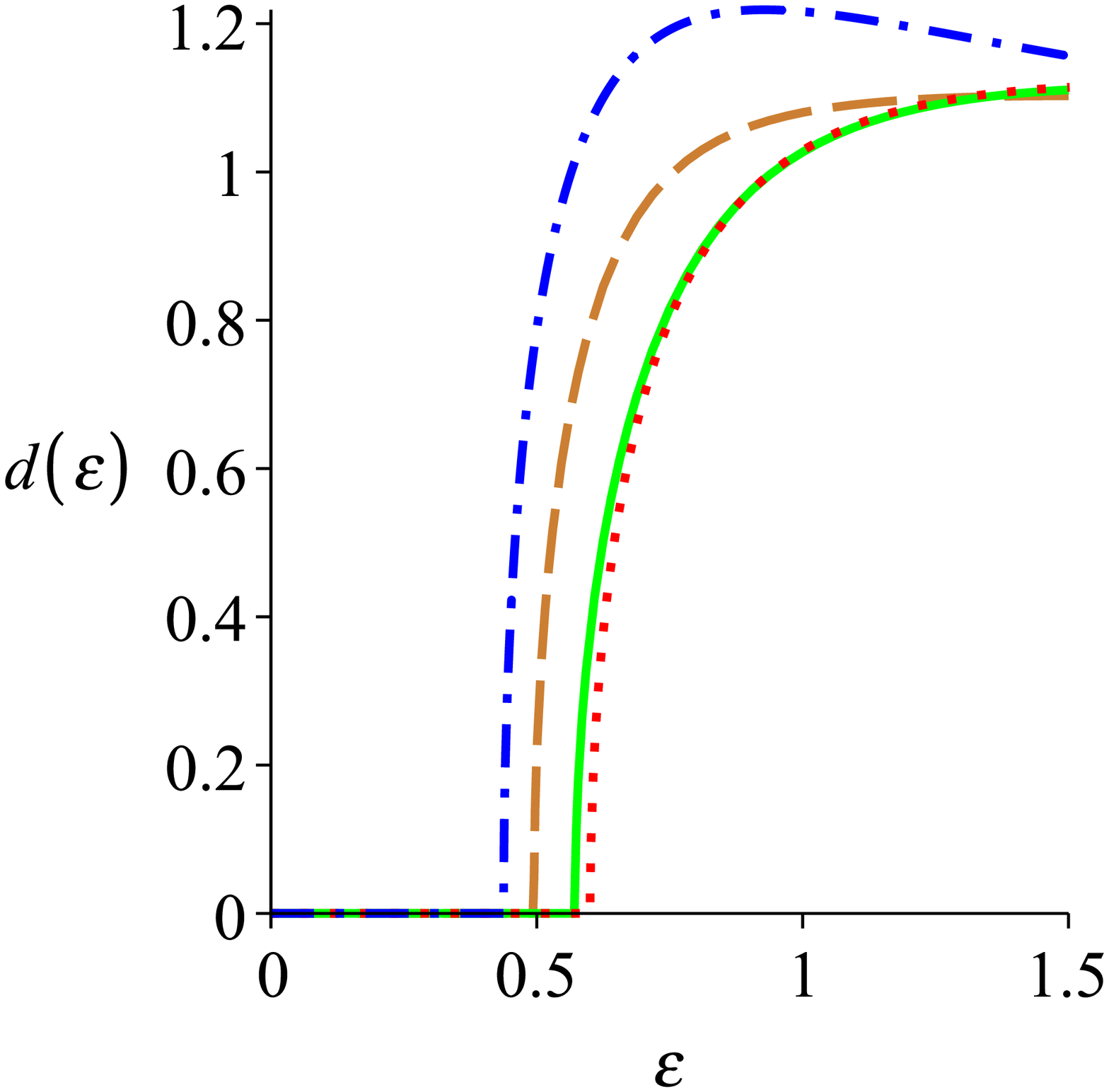}}
 \subfigure{\includegraphics[width=0.28\textwidth]{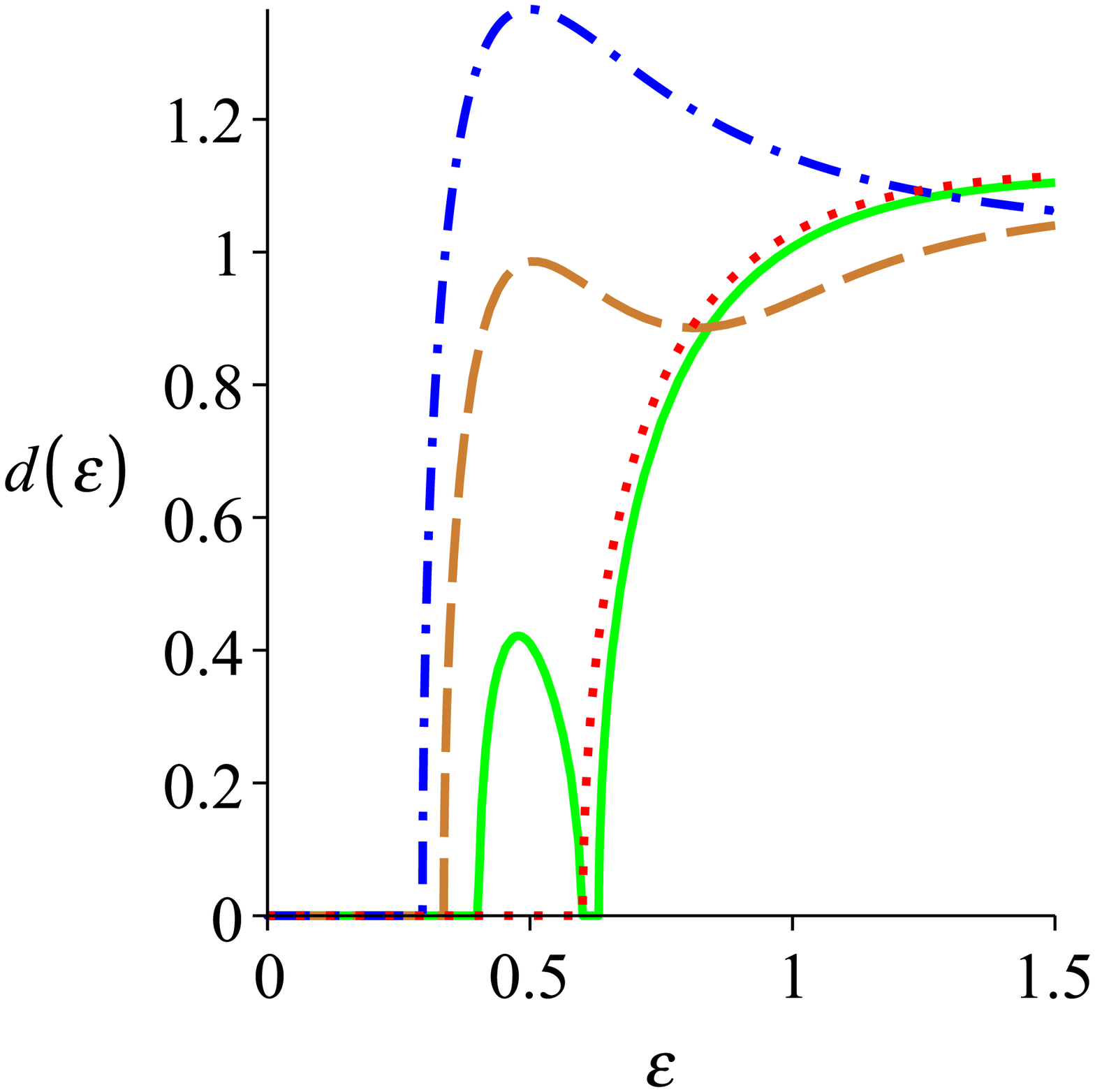}}
 \subfigure{\includegraphics[width=0.28\textwidth]{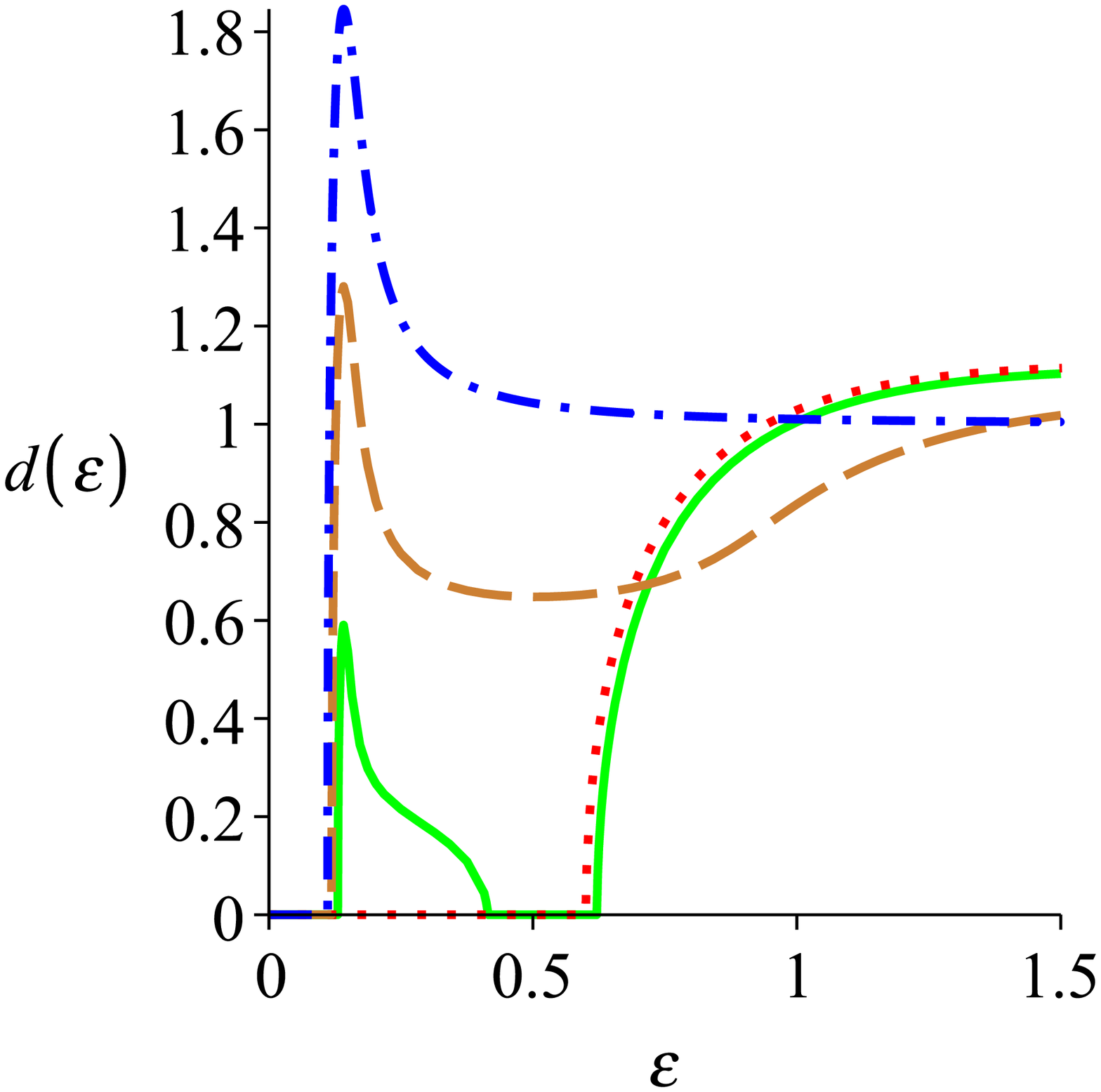}}
 \put(-430,0){(a)}
 \put(-282,0){(b)}
 \put(-134,0){(c)}
 \caption{\label{unequalphase}Dependence of the density of states of an Andreev
billiard on the difference $y=\rbr{N_1-N_2}/N$ in size of the leads with $y=0$
(dashed dotted line), $y=4/5$ (dashed line), $y=\sqrt{24}/5$ (solid line) and
$y=1$ (dotted line).  (a) At phase difference $\phi=2\pi/3$, (b) at
$\phi=5\pi/6$ and (c) at $\phi=21\pi/22$.}
\end{figure*}

\begin{figure}
 \includegraphics[width=0.6\columnwidth]{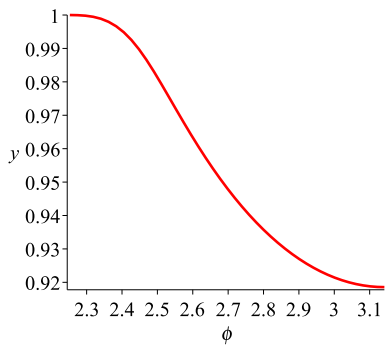}
\caption{\label{criticalunequalleads}Critical value of the difference in the
lead sizes $y$ as a function of the phase difference $\phi$ between the two
leads above which a second gap appears in the density of states.}
\end{figure}

Removing the restriction that the leads have equal size we return to a sixth
order polynomial for $g$ and $G$ when substituting \eref{gghatphaseeqn} into
\eref{phasegeqn} and then \eref{Gphaseeqn}.  Expanding $G$ as a power series in
$r$ via $G=
\sum r^{n-1}C(\en,\phi,n)$ now gives three starting values for $C(\en,\phi,1)$
and we choose the one that coincides with the result from the semiclassical
diagrams, namely $\beta\beta^{*}/\rbr{1-\rmi\en}$.  Choosing the variable $y$ to
represent the relative difference in the lead sizes
\begin{equation}
y=\frac{N_1-N_2}{N}, \qquad \beta=\cos\rbr{\frac{\phi}{2}}+\rmi
y\sin\rbr{\frac{\phi}{2}} ,
\end{equation}
leads to a particularly compact solution, and as before, we can go through our
roundabout route of finding the generating function of interest
$H(\en,\phi,y,r)$, which is recorded as \eref{unequalphaseHeqn} in the appendix.
 Although it also was not possible to verify (other than at a large number of
parameter values) this sixth order equation, from it we can obtain the
polynomial satisfied by $W(\en,\phi,y)$:
\begin{eqnarray}
&&\left[\en^2{\hat{\beta}}^2W^4+4\en
{\hat{\beta}}^2W^3+\rbr{4{\hat{\beta}}^2-\en^2+2\en^2{\hat{\beta}}^2}W^2 \right.
\nonumber\\
&& \qquad \left. {} +4\en{\hat{\beta}}^2
W-\en^2+\en^2{\hat{\beta}}^2\right]\rbr{2+\en W}^{2} \nonumber \\
&& {} + 4\en^{2}y^{2}\rbr{1-{\hat{\beta}}^2}=0 ,
 \label{semiunequalphase}
\end{eqnarray}
where we have defined $\hat{\beta}=\cos(\phi/2)$ as the real part of $\beta$
(which is equal to $\beta$ when the leads have equal size) and the evenness in
$y$ follows from the symmetry under swapping the leads and $\phi$ to $-\phi$. 
The term in the square brackets is simply \eref{semiphase} and so we recover the
result with equal leads when $y=0$.  Likewise we can check that when we only
have a single lead ($y=\pm1$) we recover a factor corresponding to
\eref{rmtsimplest} so that the phase, as expected, no longer plays a role.  From
this equation we can plot the density of states as in \fref{unequalphase} and
see how the difference in lead sizes $y$ interpolates between the result with
equal leads above and the density of states with a single lead in
\eref{singledossol}.  Note in particular that the peak in the density of states
as the phase difference nears $\pi$ vanishes slowly as $y$ approaches $\pm1$ so
that we can see a second gap appear in the density of states for leads differing
distinctly in channel numbers (for example, see the solid line in
\frefs{unequalphase}b and c).  Numerically we can extract the critical value of
$y$ for each phase difference $\phi$ above which we see a second gap.  We plot
this in \fref{criticalunequalleads} where we see that the second gap only
appears for particularly unequal leads and at reasonable phase differences.

\section{\label{ehrenfest}Ehrenfest time dependence}

So far we have been looking at the regime where the Ehrenfest time $\tE \sim
\vert\ln\hbar\vert$, the time below which wave packets propagate essentially
classically (and above which wave interference dominates), is small compared to
the dwell time $\tD$, the typical time the trajectories spend inside the
scattering region.  This is the same limit described by RMT and we have seen the
agreement between semiclassics and RMT in \srefs{singlelead} and~\ref{twoleads}
above.  Moving away from this limit we can treat the typical effect of the
Ehrenfest time on the correlation functions $C(\en,n)$, for now for the simplest
case of a single lead and no magnetic field.  To contribute in the semiclassical
limit, the correlated trajectories should have an action difference of the order
of $\hbar$ which in turn means that the encounters have a duration of the order
of the Ehrenfest time.  Increasing this relative to the dwell time, or
increasing the ratio $\tau=\tE/\tD$, then increases the possibility that all the
trajectories travel together for their whole length in a correlated band. 
Likewise the probability of forming the diagrams (as in
\fref{fourtrajectorystructures}) considered before reduces.  All told, the
Ehrenfest time dependence~\cite{waltneretal10} leads to the simple replacement
\begin{equation}
 C(\en,\tau,n)=C(\en,n)\rme^{-(1-\rmi n\en)\tau}+\frac{1-\rme^{-(1-\rmi
n\en)\tau}}{1-\rmi n\en} .
 \label{ehrenfestcorr}
\end{equation}
This replacement leaves the $n=1$ term unchanged and had previously been shown
for $n=2$ in \ocite{wj06} and $n=3$ in \ocite{br06b}.  The exponential growth of
differences between trajectories due to the chaotic motion means that we just
add the first term from the previous diagrams with encounters in
\eref{ehrenfestcorr} to the second term from the bands as their opposing length
restrictions lead to a negligible overlap.  In fact this separation into two
terms was shown~\cite{wj05,jw06} to be a direct consequence of the splitting of
the classical phase space into two virtually independent subsystems.

\begin{figure*}
 \centering
\subfigure{\includegraphics[width=0.42\textwidth]{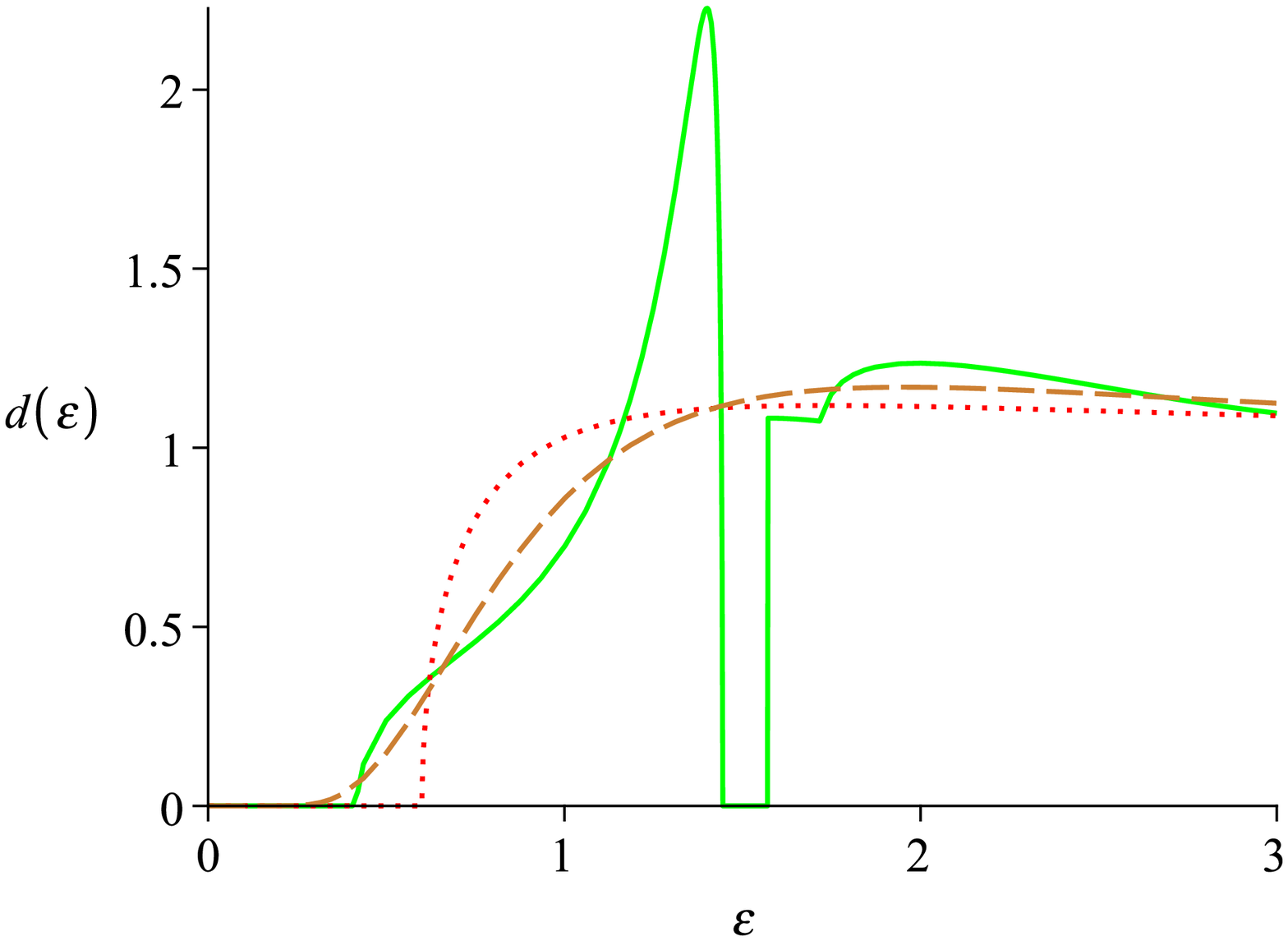}}
\subfigure{\includegraphics[width=0.42\textwidth]{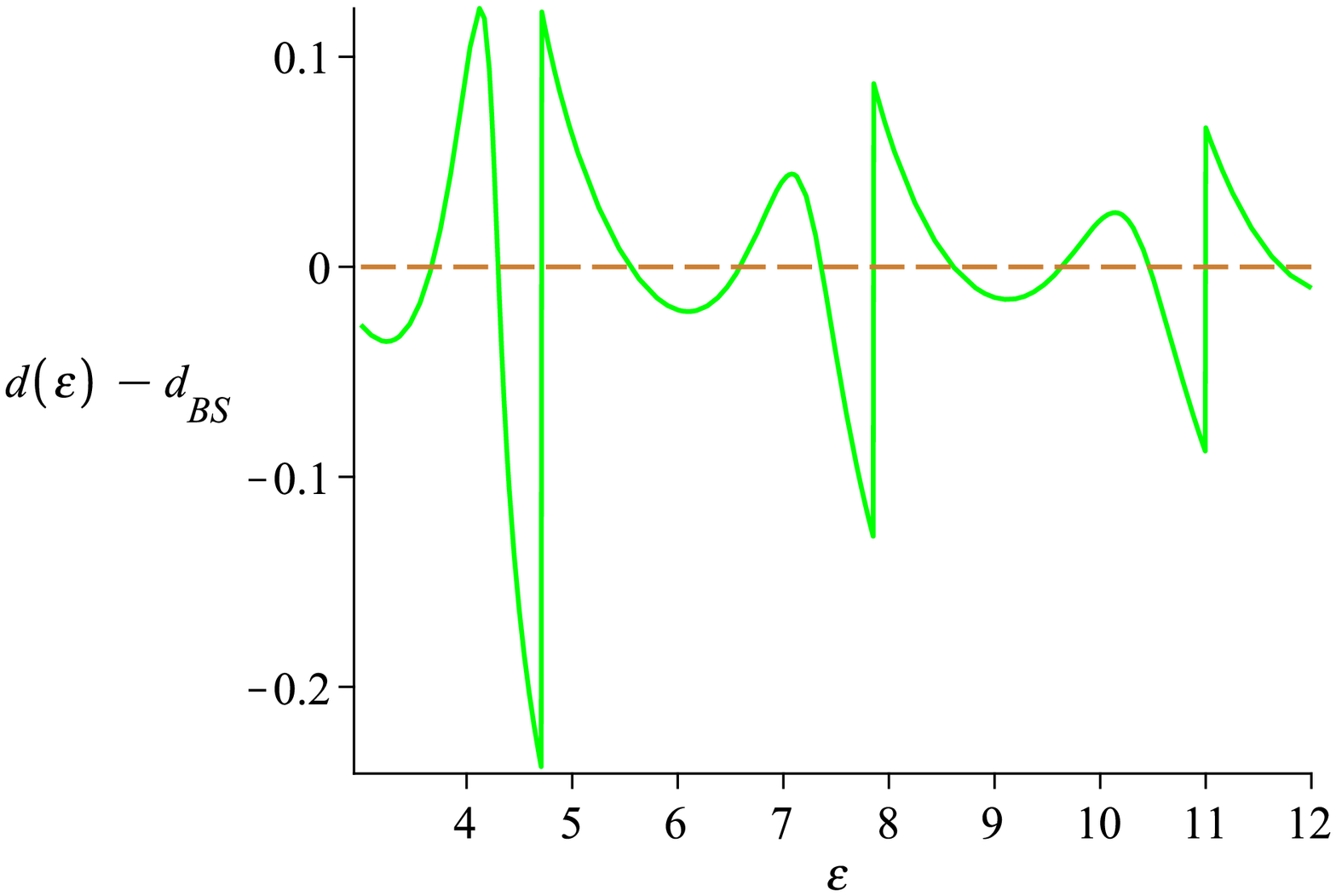}}
 \put(-426,0){(a)}
 \put(-210,0){(b)}
 \caption{\label{ehrenfestsingleplot}(a) Density of states for $\tau=\tE/\tD=2$
(solid line), along with the BS (dashed) limit $\tau\to\infty$ and the RMT
(dotted) limit $\tau=0$, showing a second gap just below $\en\tau=\pi$. (b)
Ehrenfest time related $2\pi/\tau$-periodic oscillations in the density of
states after subtracting the BS curve.}
\end{figure*}

We leave the technical demonstration of \eref{ehrenfestcorr} to
\ocite{waltneretal10} but the result follows by treating the diagrams considered
before, which are created by sliding encounters together or into the lead (like
the process depicted in \frefs{fourtrajectorystructures}
and~\ref{fourtrajectorystructureleads}), as part of a continuous deformation of
a single diagram.  With a suitable partition of this family one can see that
each set has the same $\tE$ dependence and hence that \eref{ehrenfestcorr} holds
for all $n$.  It is clear that in the limit $\tau=0$ \eref{ehrenfestcorr}
reduces to the previous (and hence RMT) results while in the opposite limit,
$\tau=\infty$, substituting \eref{ehrenfestcorr} into \eref{dossingle} and
performing a Poisson summation we obtain the Bohr-Sommerfeld (BS)~\cite{sb99}
result
\begin{equation}
 d_\mathrm{BS}(\en)=\rbr{\frac{\pi}{\en}}^2\frac{\cosh(\pi/\en)}{
\sinh^2(\pi/\en)} .
\label{bohrsommerfeldeqn}
\end{equation}
This result was previously found semiclassically by \ocite{ihraetal01} and
corresponds to the classical limit of bands of correlated trajectories.

For arbitrary Ehrenfest time dependence we simply substitute the two terms in
\eref{ehrenfestcorr} into \eref{dossingle}.  With the second term we include
$1-(1+\tau)\rme^{-\tau}$ from the constant term (this turns out to simplify the
expressions) and again perform a Poisson summation to obtain
\begin{eqnarray}
\label{ehrenfestdostwo} d_2(\en,\tau)&=&1-(1+\tau)\rme^{-\tau}  \\ 
&&
{}+2\mathrm{Im}\sum_{n=1}^{\infty}\frac{(-1)^n}{n}\pdiff{}{\en}\rbr{\frac{
1-\rme^{-(1-\rmi n\en)\tau}}{1-\rmi n\en}}\nonumber\\
  &=&d_\mathrm{BS}(\en) \nonumber \\
&& {} -\exp\rbr{-\frac{2\pi
k}{\en}}\rbr{d_\mathrm{BS}(\en)+\frac{2k(\pi/\en)^2}{\mathrm{sinh}(\pi/\en)}} ,
\nonumber
\end{eqnarray}
where $k=\lfloor(\en\tau+\pi)/(2\pi)\rfloor$ involves the floor function, and we
see that this function is zero for $\en\tau<\pi$.

\begin{figure}
 \centering
 \subfigure{\includegraphics[width=0.49\columnwidth]{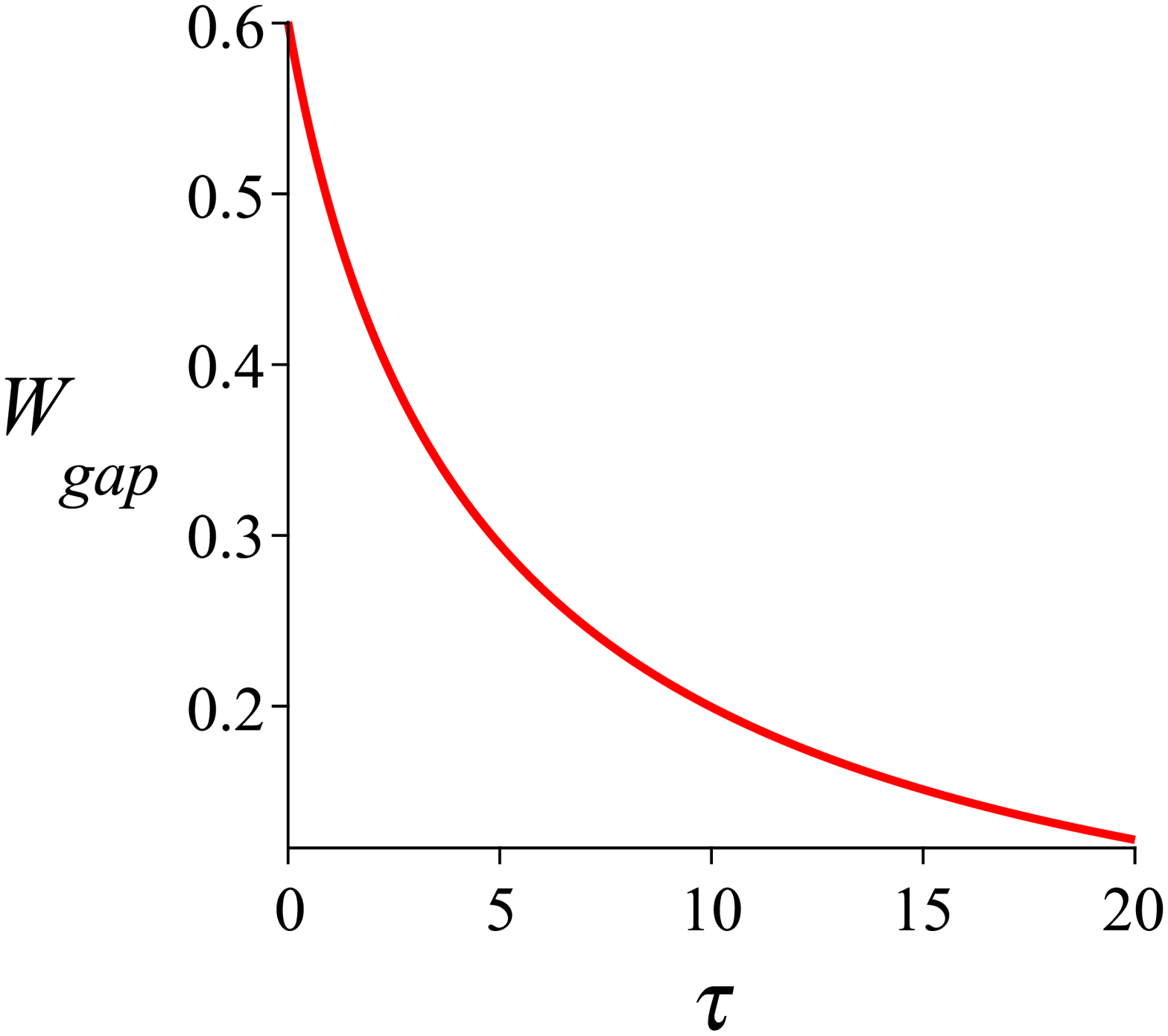}}
\subfigure{\includegraphics[width=0.49\columnwidth]{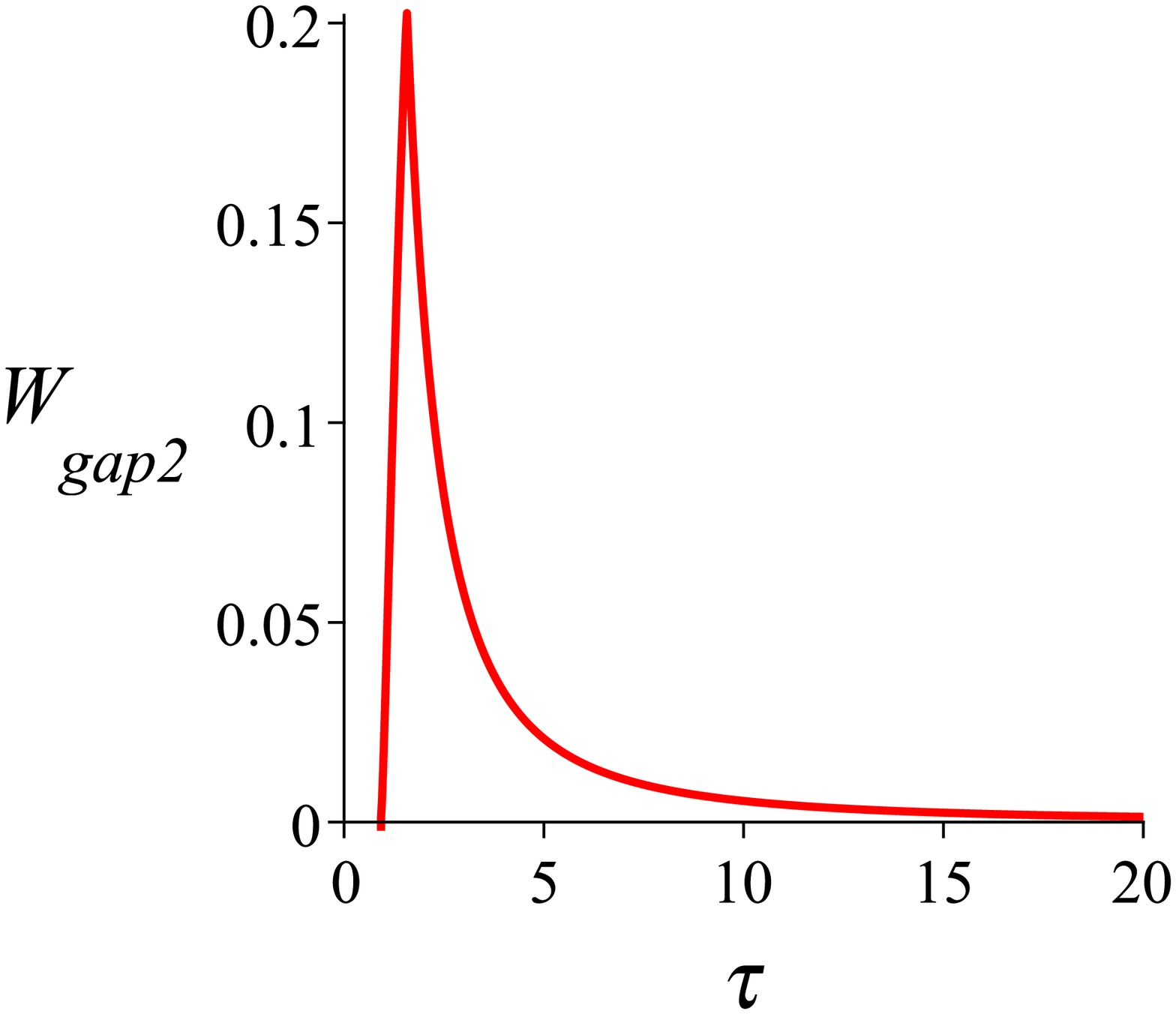}}
 \put(-240,0){(a)}
 \put(-115,0){(b)}
 \caption{\label{dosgapwidth}(a) Width (and end point) of the first gap and (b)
width of the second gap as a function of $\tau$.}
\end{figure}
\begin{figure*}
 \subfigure{\includegraphics[width=0.42\textwidth]{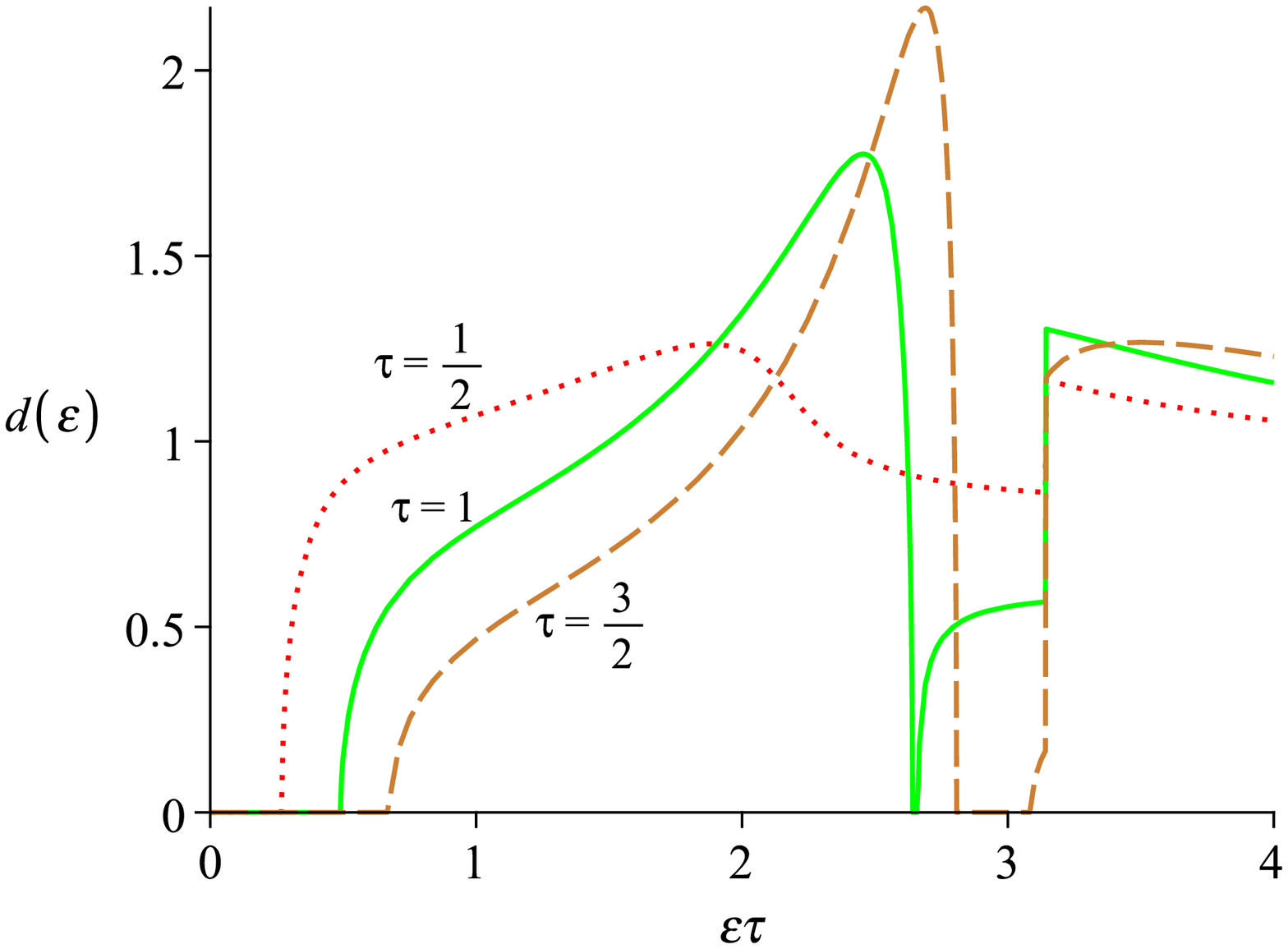}}
 \subfigure{\includegraphics[width=0.42\textwidth]{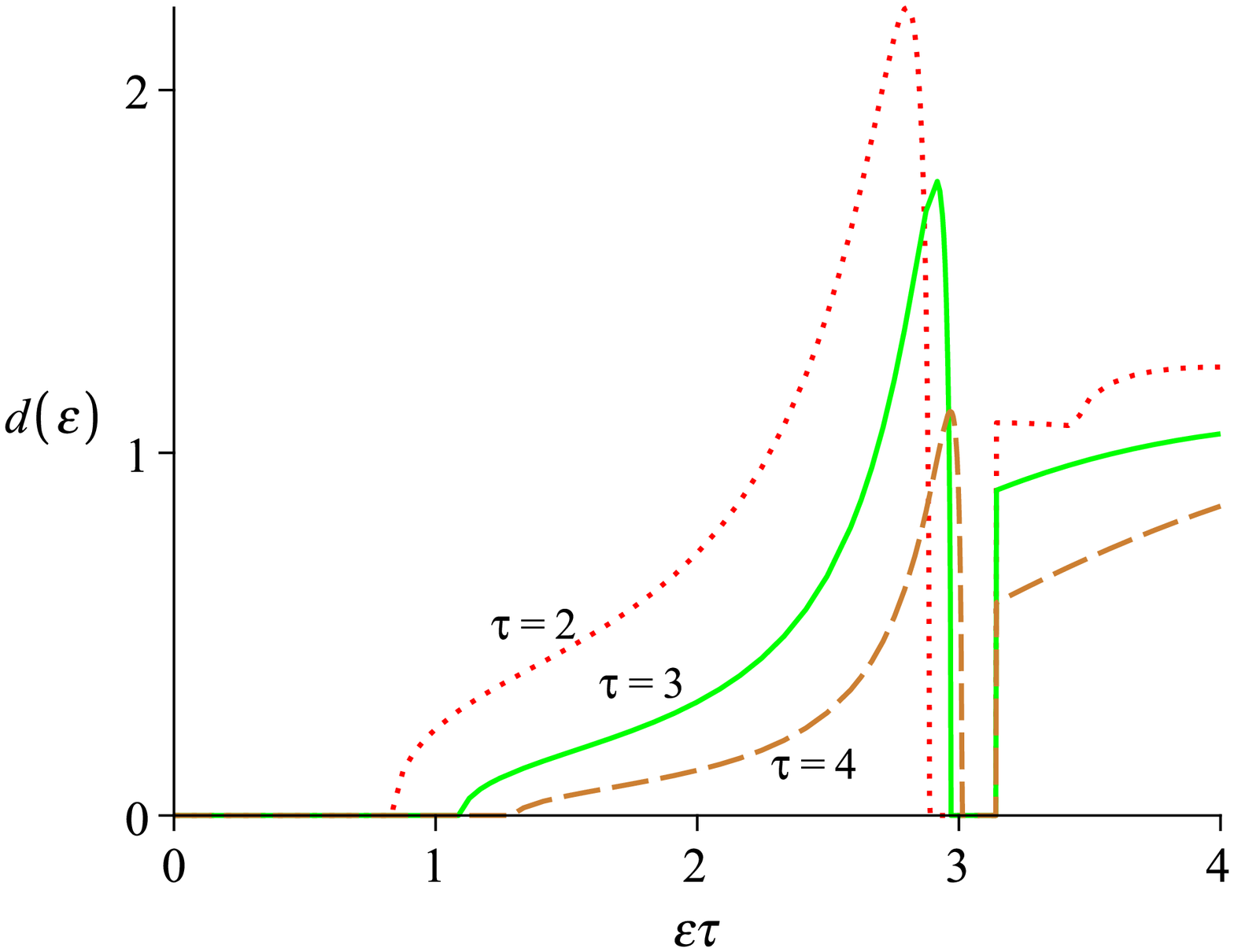}}
 \put(-426,0){(a)}
 \put(-210,0){(b)}
 \put(-180,95){\includegraphics[width=0.2\textwidth]{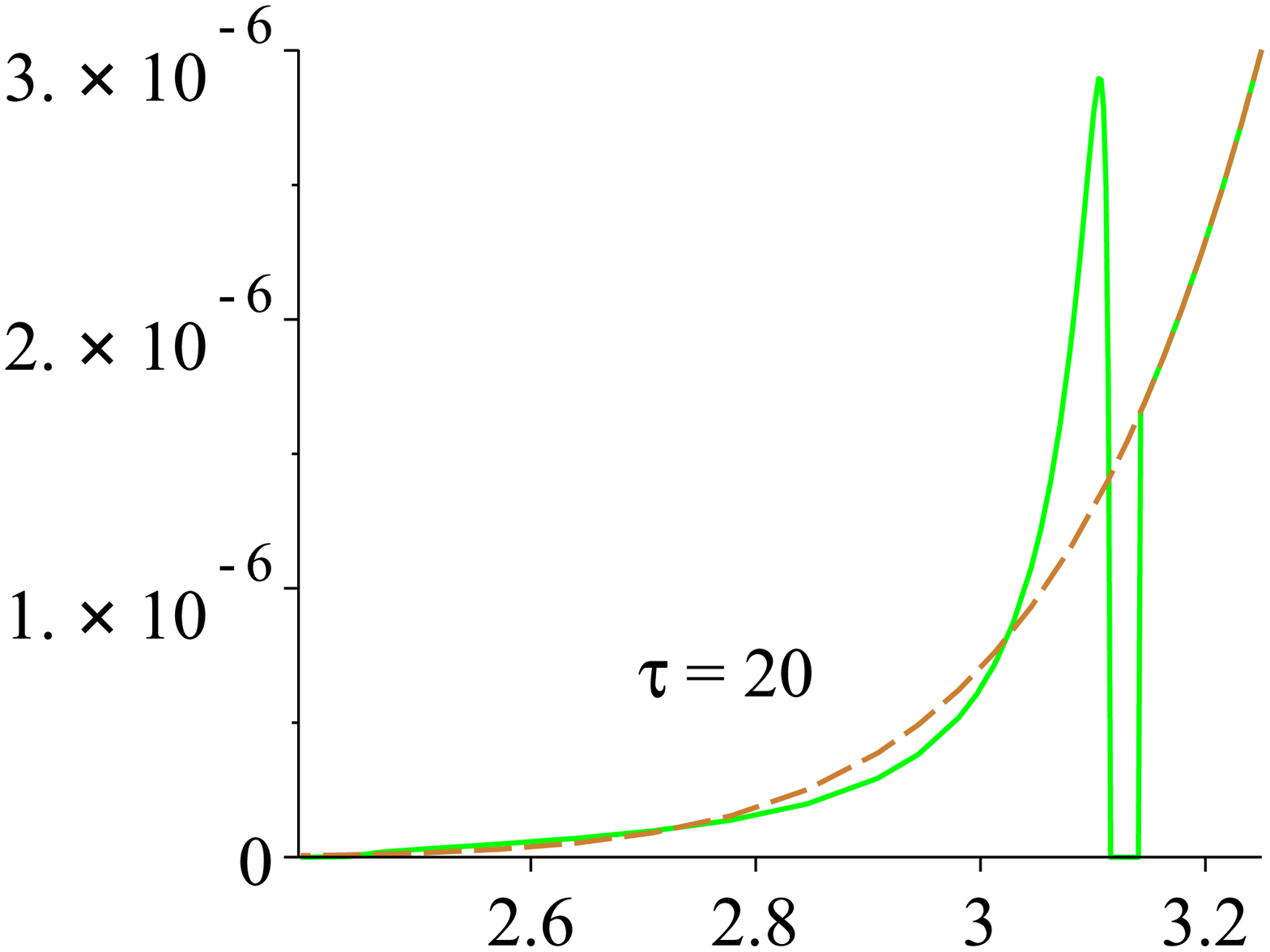}}
 \caption{\label{dosdiffehrenfest}Density of states as a function of
$\en\tau=E/E_\mathrm{E}$ for various values of $\tau$ showing the appearance of
a second gap below $\en\tau=\pi$. Inset: Density of states for $\tau=20$ (solid
line) together with the BS limit (dashed).}
\end{figure*}

Of course the first term in \eref{ehrenfestcorr} also contributes and when we
substitute into \eref{dossingle} we obtain two further terms from the energy
differential.  These however may be written, using our semiclassical generating
functions, as
\begin{eqnarray}
\label{ehrenfestdosone}
d_1(\en,\tau)&=&\rme^{-\tau}\cbr{1-2\mathrm{Re}\;\rme^{\rmi\en\tau}H(\en,-\rme^{
\rmi\en\tau})} \\
& &
{}+\tau\rme^{-\tau}\cbr{1-2\mathrm{Re}\;\rme^{\rmi\en\tau}G(\en,-\rme^{
\rmi\en\tau})} . \nonumber 
\end{eqnarray}
Because $G$ and $H$ are given by cubic equations, we can write this result
explicitly as
\begin{eqnarray}
\label{ehrenfestdosoneexplicit}
d_1(\en,\tau) &=& \frac{\sqrt{3}\rme^{-\tau}}{6\en} \mathrm{Re}
\cbr{Q_{+}(\en,\tau)-Q_{-}(\en,\tau)} \\
&& {} + \frac{\sqrt{3}\tau\rme^{-\tau}}{6} \mathrm{Re}
\cbr{P_{+}(\en,\tau)-P_{-}(\en,\tau)} , \nonumber
\end{eqnarray}
where
\begin{eqnarray}
Q_{\pm}(\en,\tau)&=&\left[8-\frac{24\en\rbr{1-\cos(\en\tau)}}{\sin(\en\tau)}
-24\en^{2} \right.\nonumber \\ 
&& \quad {} -\frac{24\en^{2}\rbr{1-\cos(\en\tau)}}{\sin^{2}(\en\tau)}
+\frac{6\en^{3}\rbr{1-\cos(\en\tau)}}{\sin(\en\tau)} \nonumber \\
&& \quad {}
+\frac{2\en^{3}\rbr{2-3\cos(\en\tau)+\cos^{3}(\en\tau)}}{\sin^{3}(\en\tau)}
\nonumber \\
&& \left. \quad {}
\pm\frac{6\en\sqrt{3D}\rbr{1-\cos(\en\tau)}}{\sin^{2}(\en\tau)}\right]^{\frac{1}
{3}} ,
\end{eqnarray}
\begin{eqnarray}
P_{\pm}(\en,\tau)&=&\left[\frac{36\en}{\rbr{1+\cos(\en\tau)}^{2}}-\frac{9\en^{2}
\sin(\en\tau)}{\rbr{1+\cos(\en\tau)}^{3}} \right.  \\  && \left. \quad {}
+\frac{\en^3}{\rbr{1+\cos(\en\tau)}^{3}}\pm\frac{3\sqrt{3D}}{\rbr{
1+\cos(\en\tau)}^{2}}\right]^{\frac{1}{3}} . \nonumber
\end{eqnarray}
These all involve the same discriminant $D$ and so the differences in
\eref{ehrenfestdosoneexplicit} are only real (and hence $d_1(\en,\tau)$ itself
is non-zero) when
\begin{eqnarray}
D(\en,\tau)&=&\en^4-8\en^3\sin(\en\tau)+4\en^2\cbr{5+6\cos(\en\tau)} \nonumber
\\
&& {} +24\en\sin(\en\tau)-8\cbr{1+\cos(\en\tau)} ,
 \label{discriminant}
\end{eqnarray}
is positive. Recalling that the second contribution is zero up to $\en\tau=\pi$,
the complete density of states is therefore zero up to the first root of
$D(\en,\tau)$.  The width of this gap is then solely determined by the
contribution from quantum interference terms given by the trajectories with
encounters.  The hard gap up to the first root shrinks as $\tau$ increases (see
\fref{dosgapwidth}a) and when taking the limit $\tau\rightarrow\infty$ while
keeping the product $\en\tau$ constant \eref{discriminant} reduces to
$-8\cbr{1+\cos(\en\tau)}$ which has its first root at $\en\tau=\pi$.  The gap
then approaches $E=\pi E_\mathrm{E}$ for $\tau\gg1$ where
$E_{\mathrm{E}}=\hbar/2\tE$ is the Ehrenfest energy.  So one indeed observes a
hard gap up to $\pi E_\mathrm{E}$ in the limit $\tau\rightarrow\infty$ at fixed
$\en\tau$ in agreement with the quasiclassical result of \ocite{ma09}.  

Alongside this reduction in size of the first gap, which was predicted by
effective RMT~\cite{beenakker05}, when $\tau\geq0.916$ the discriminant
\eref{discriminant} has additional roots. Between the second and third root
$D(\en,\tau)$ is also negative and a second gap appears.  As $\tau$ increases
the roots spread apart so the gap widens.  For example, the complete density of
states for $\tau=2$ is shown in \fref{ehrenfestsingleplot}a along with the
oscillatory behavior visible at larger energies (with period $2\pi/\tau$) in
\fref{ehrenfestsingleplot}b.  There the second gap is clearly visible and only
ends when the second contribution $d_2(\en,\tau)$ becomes non-zero at
$\en\tau=\pi$.  In fact for $\tau>\pi/2$ the third root of $D(\en,\tau)$ is
beyond $\en\tau=\pi$ so the second gap is cut short by the jump in the
contribution $d_2(\en,\tau)$.  Since the second root also increases with
increasing $\tau$ the gap shrinks again, as can be seen in \fref{dosgapwidth}b.

To illustrate this behavior further, the density of states is shown for
different values of $\tau$ in \fref{dosdiffehrenfest}. One can see first the
formation and then the shrinking of the second gap.  As can be seen in the inset
of \fref{dosdiffehrenfest}b the second gap persists even for large values of
$\tau$ and the size of the first hard gap converges slowly to $\en\tau=\pi$. The
plot for $\tau=20$ also shows how the density of states converges to the BS
result.

\subsection{Effective RMT}

\begin{figure*}
\centering
 \subfigure{\includegraphics[width=0.28\textwidth]{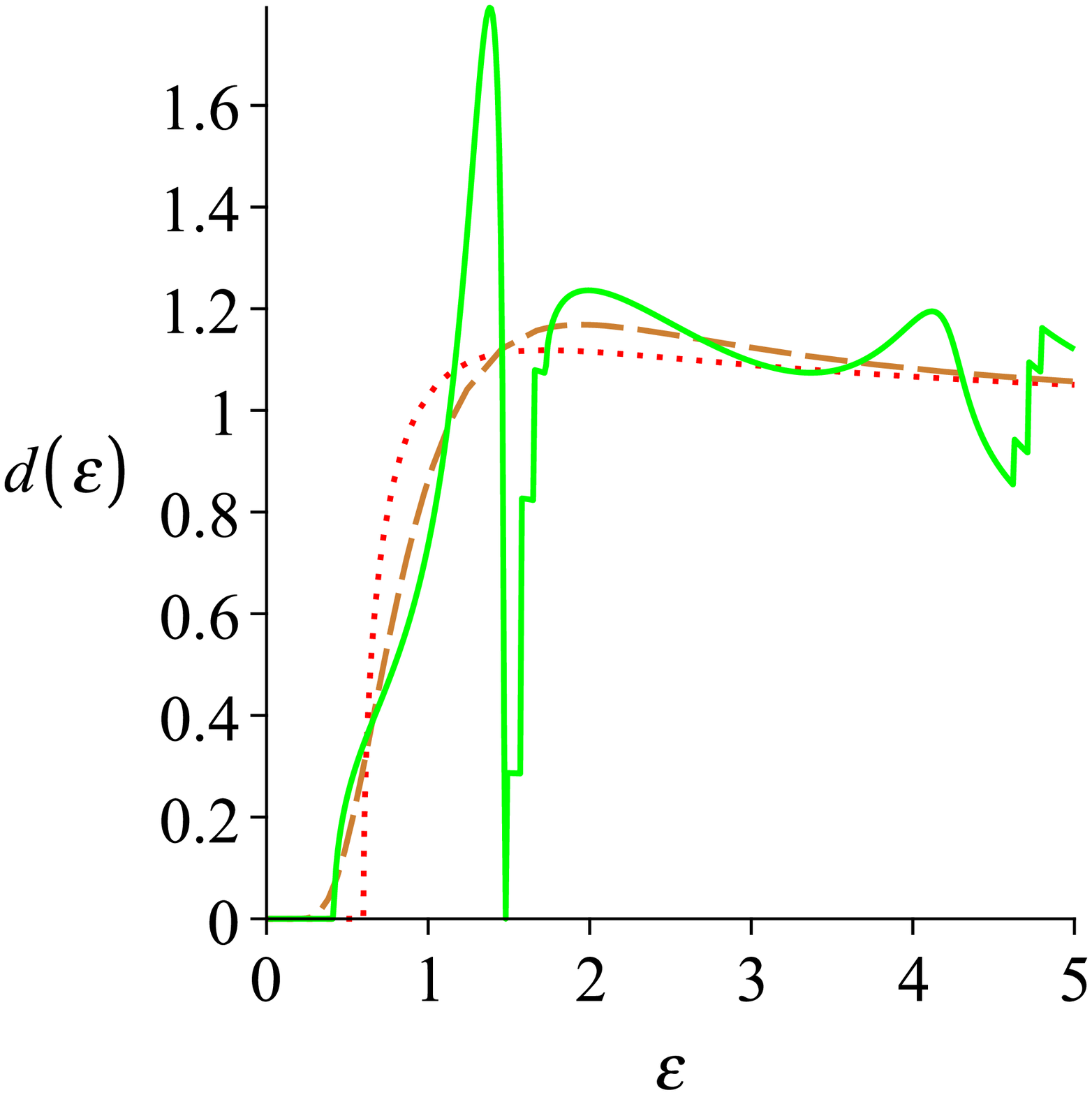}}
 \subfigure{\includegraphics[width=0.28\textwidth]{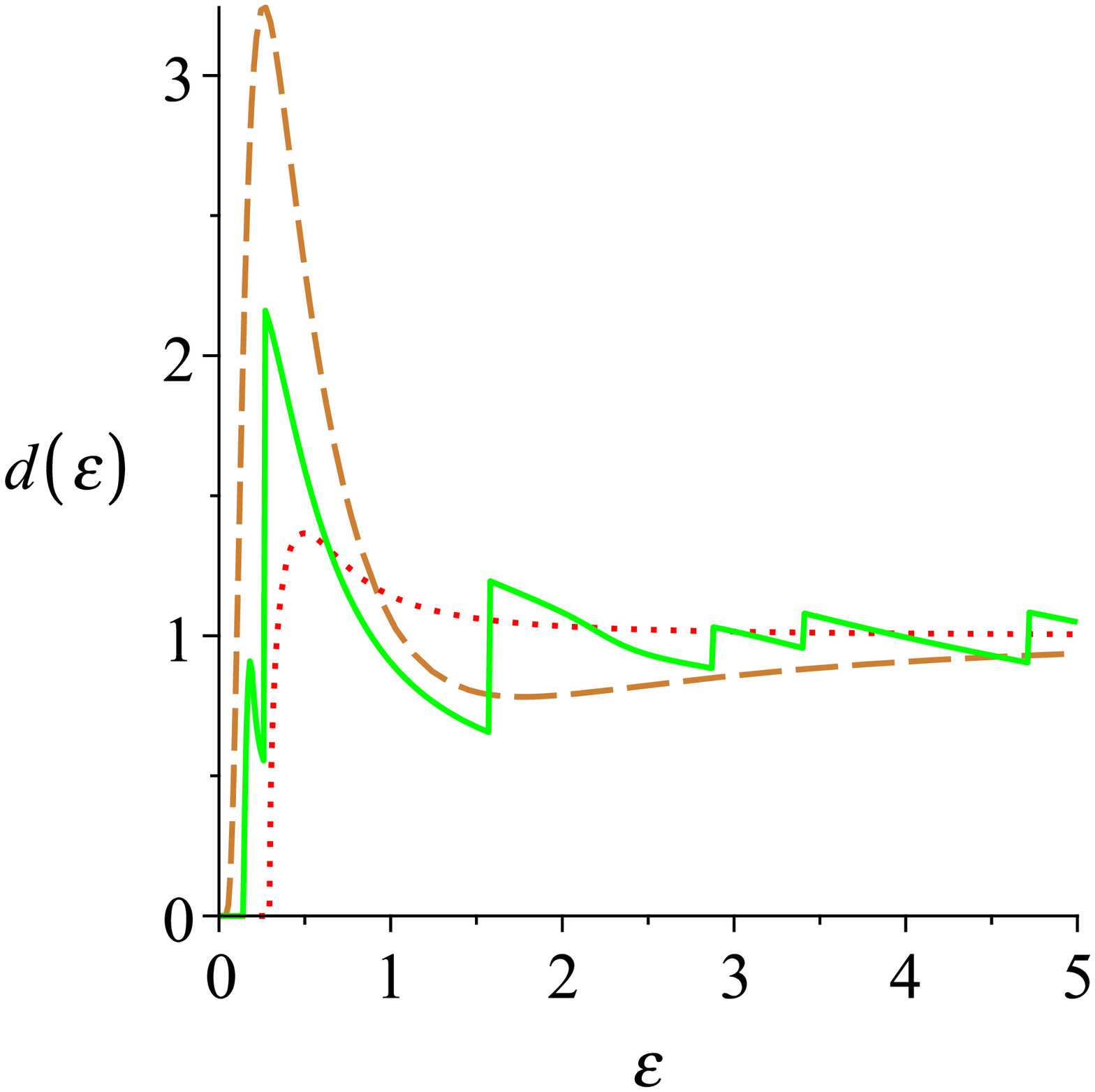}}
 \subfigure{\includegraphics[width=0.28\textwidth]{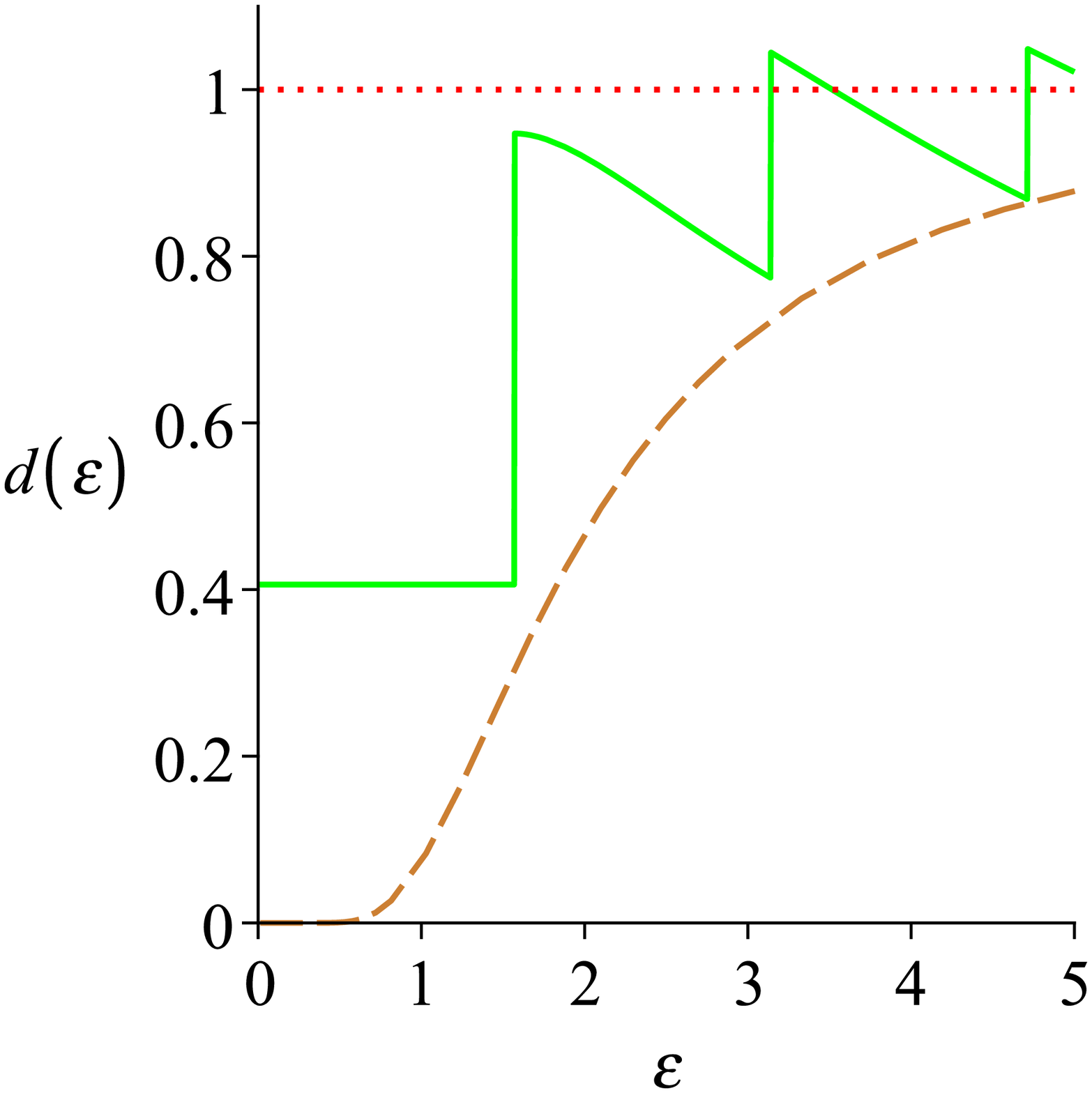}}
 \put(-430,0){(a)}
 \put(-282,0){(b)}
 \put(-134,0){(c)}
 \caption{\label{ehrenfestphase}Density of states for $\tau=2$ (solid line)
along with the $\tau=0$ (dotted) and $\tau=\infty$ (dashed) limits for a chaotic
Andreev billiard with phase difference (a) $\phi=\pi/18$, (b) $\phi=5\pi/6$ and
(c) $\phi=\pi$.}
\end{figure*}

As mentioned above, the shrinking of the first gap has been predicted by
effective RMT where the effect of the Ehrenfest time is mimicked~\cite{sgb03} by
reducing the number of channels in the lead by a factor $\rme^{\tau}$ (to
correspond to the part of classical phase space where the trajectories are
longer than the Ehrenfest time) and to multiply the scattering matrix by the
phase $\rme^{\rmi\en\tau/2}$ to represent the energy dependence of the lead. 
The random matrix diagrammatic averaging leads to the set of
equations~\cite{beenakker05,gjb05}
\begin{eqnarray}
\label{effrmteqn}
W^2+1&=&W_{2}^2 \\
W+W_{2}\sin u &=& -\frac{\en}{2}W_{2}\rbr{W_{2}+\cos u +W \sin u} , \nonumber
\end{eqnarray}
where $u=\en\tau/2$ and the density of states is given by (for $u<\pi/2$)
\begin{equation}
d(\en,\tau)=-\rme^{-\tau}\mathrm{Im} \rbr{W-\frac{u}{\cos u}W_{2}} .
\label{effrmtdos}
\end{equation}
The equations in \eref{effrmteqn} can be simplified to obtain a cubic for $W$
(and $W_2$) and in this form we can compare with our semiclassical results.  In
fact, making the substitution $H=\cbr{\rmi W-1}/2r$ and setting
$r=-\exp(\rmi\en\tau)$ to get the first part in \eref{ehrenfestdosone} in the
form of the first term in \eref{effrmtdos} we obtain exactly the same polynomial
and hence agreement.  Likewise when we substitute $G=-\cbr{\rmi u W_{2}/\cos
u+\tau}/2r\tau$ we obtain the same polynomial for the second part, albeit with
the real offset $\tan u$ which does not affect the density of states.  

Of course this agreement provides semiclassical support for the phenomenological
approach of effective RMT as well as showing that \eref{effrmtdos} is valid for
$u$ beyond $\pi/2$.  In principle then the second gap could also have been found
using effective RMT, but of course effective RMT misses the second contribution
to the density of states $d_{2}(\en,\tau)$.  This contribution turns out to be
straightforward to obtain semiclassically, and can be compared to the bands
treated in \ocite{ma09}.

\subsection{\label{ehrenfesttwo}Two superconducting leads}

If we include the effect of a symmetry breaking magnetic field then, because of
the way this affects the contribution of different sized encounters (as
described in \sref{singlemag}), such a simple replacement as in
\eref{ehrenfestcorr} no longer holds.  This situation has however been treated
using effective RMT~\cite{gjb05} allowing them to also determine how the
critical magnetic field (at which the gap closes) depends on the Ehrenfest time.
 Without the simple replacement of the magnetic field dependent correlations
functions we are currently unable to confirm this result semiclassically.  But
if we look at two superconducting leads (for simplicity of equal size) at
different phase this effect can be included in the channel sum and treated as
above (the effective RMT result can be found by a simple modification of the
treatment in \ocite{gjb05}).  Important to remember is that the second part (of
\eref{ehrenfestcorr}) corresponds to bands of trajectories that are correlated
for their whole length and so they all start and end together (in the same
leads). Therefore the second contribution has to be multiplied by
$[1+\cos(n\phi)]/2$ leading to
\begin{eqnarray}\label{ehrenfestwocorr}
 C(\en,\phi,\tau,n)&=&C(\en,\phi,n)\rme^{-(1-\rmi n\en)\tau} \\
&& {} +\frac{1+\cos(n\phi)}{2}\frac{1-\rme^{-(1-\rmi n\en)\tau}}{1-\rmi n\en} .
\nonumber
\end{eqnarray}
The first part of the density of states for non zero Ehrenfest time then remains
as in \eref{ehrenfestdosone} but with $G(\en,r)$ and $H(\en,r)$ replaced by
$G(\en,\phi,r)$ and $H(\en,\phi,r)$, respectively. The second contribution in
this case however may be written as the average of the $\phi=0$ contribution and
a contribution with the full phase difference $\phi$,
\begin{equation}
 d_2(\en,\phi,\tau)=\frac{1}{2}\cbr{d_2^\prime(\en,0,\tau)+d_2^\prime(\en,\phi,
\tau)} .
 \label{ehrenfesttwosecond}
\end{equation}
Here $d_2^\prime(\en,\phi,\tau)$ may be again written as the sum of the
$\tau=\infty$ result
\begin{eqnarray}
 \label{bstwo}
&& d_2^{\prime(1)}(\en,\phi,\tau)=\frac{\pi}{2\en^2\sinh^2\rbr{\pi/\en}}  \\
 && \times \left[\rbr{\pi+2\pi k_1-\phi}\cosh\rbr{\frac{\pi-2\pi
k_1+\phi}{\en}}\right. \nonumber \\
 && \qquad \left. {} +\rbr{\pi-2\pi k_1+\phi}\cosh\rbr{\frac{\pi+2\pi
k_1-\phi}{\en}}\right] , \nonumber
\end{eqnarray}
and some correction
\begin{eqnarray}
\label{d2two}
&& d_2^{\prime(2)}(\en,\phi,\tau)=-\frac{\pi}{2\en^2\sinh^2\rbr{\pi/\en}} \\ 
&& {} \times \left\{\cbr{\pi\cosh\rbr{\frac{\pi}{\en}}+\rbr{2\pi
k_2-\phi}\sinh\rbr{\frac{\pi}{\en}}}\rme^{-\frac{2\pi k_2-\phi}{\en}}\right.
\nonumber \\
&& \left. {} +\cbr{\pi\cosh\rbr{\frac{\pi}{\en}}+\rbr{2\pi
k_3+\phi}\sinh\rbr{\frac{\pi}{\en}}}\rme^{-\frac{2\pi k_3+\phi}{\en}}\right\} ,
\nonumber
\end{eqnarray}
with $k_1=\lfloor\rbr{\pi+\phi}/(2\pi)\rfloor$,
$k_2=\lfloor\rbr{\en\tau+\pi+\phi}/(2\pi)\rfloor$ and
$k_3=\lfloor\rbr{\en\tau+\pi-\phi}/(2\pi)\rfloor$. Since the $k_i$ and $\phi$
only occur in the combinations $2\pi k_1-\phi$, $2\pi k_2-\phi$ and $2\pi
k_3+\phi$ it is obvious that these contributions have oscillations in the phase
$\phi$ with period $2\pi$. It can also be easily seen that for $\phi=0$ the
previous result for the density of states in the Ehrenfest regime is reproduced.

\begin{figure*}
 \subfigure{\includegraphics[width=0.42\textwidth]{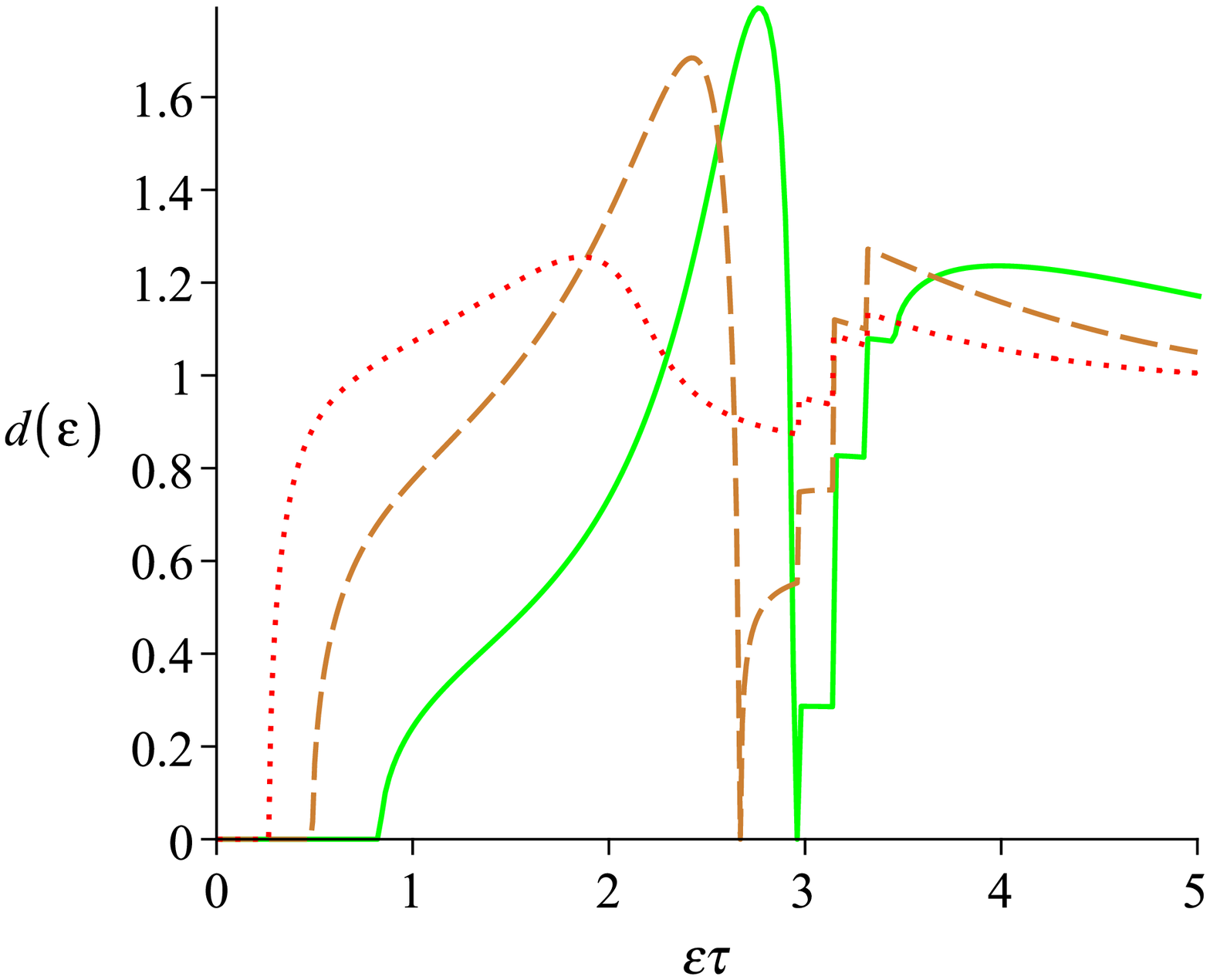}}
 \subfigure{\includegraphics[width=0.42\textwidth]{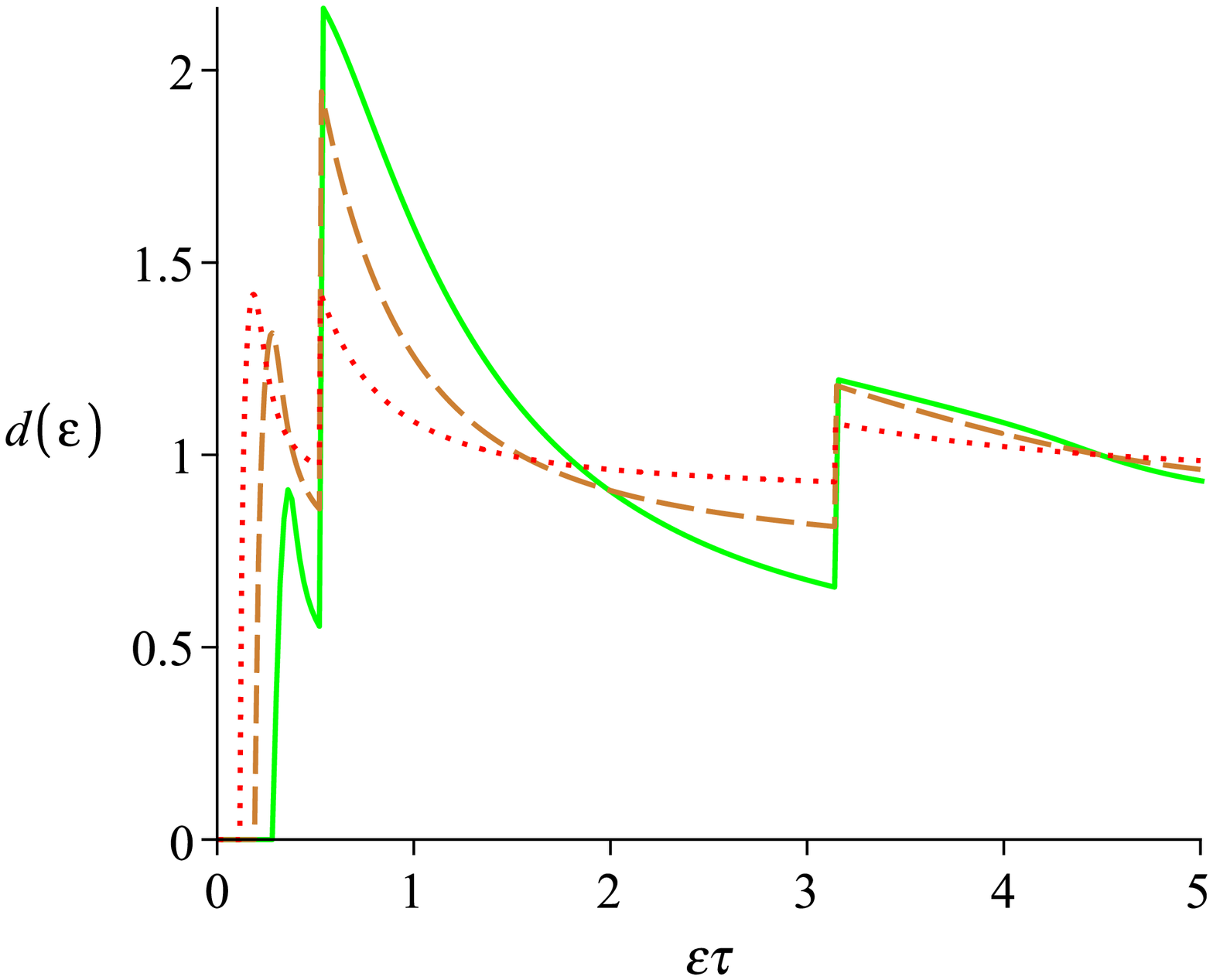}}
 \put(-426,0){(a)}
 \put(-210,0){(b)}
 \caption{\label{ehrenfesttauphase} Density of states for $\tau=1/2$ (dotted
line), $\tau=1$ (dashed) and $\tau=2$ (solid) showing the phase dependent jumps
for phase difference (a) $\phi=\pi/18$ and (b) $\phi=5\pi/6$.}
\end{figure*}

With $\vert\phi\vert<\pi$ we have $k_1=k_2=k_3=0$ for
$\en\tau<\pi-\vert\phi\vert$. Therefore one again sees that $d_2=0$ as long as
$\en\tau<\pi-\vert\phi\vert$. The first part $d_2^{\prime(1)}$ equals the
Bohr-Sommerfeld result \eref{bohrsommerfeldeqn}, so in the limit $\tau=\infty$
this result is reproduced again. The oscillations in $\en$ seen in the $\phi=0$
case which have a period of $2\pi/\tau$ can still be seen due to the fact that
the $\phi=0$ result enters $d_2(\en,\phi,\tau)$ even if $\phi\neq0$. However one
gets additional (but smaller) steps at energies satisfying
$\en=[(2m-1)\pi\mp\phi]/\tau$ for integer $m$.

We plot the density of states for $\tau=2$, along with the $\tau=0$ and
$\tau=\infty$ limits in \fref{ehrenfestphase} for different values of the phase
difference.  We can see that as the phase difference increases the second
intermediate gap (\cf \fref{ehrenfestsingleplot}a) shrinks quickly. The reason
for this shrinking is twofold: On the one hand, the gap in the RMT-like contribution
shrinks, and on the other hand, the second contribution is zero only up to
$\en\tau=\pi-\vert\phi\vert$. Moreover if $\phi\to\pi$ the modified correlation
function tends to zero so the density of states converges to
$(1+\tau)\rme^{-\tau}+d_2(\en,\tau)$. For a finer look at the Ehrenfest time
dependence and the phase dependent jumps we plot the density of states for
$\tau=1/2$, 1 and 2 for phases $\phi=\pi/18$ and $5\pi/6$ in
\fref{ehrenfesttauphase}.

\section{Conclusions}

From the semiclassical treatment of the density of states of chaotic Andreev
billiards we have seen how fine correlations between ever larger sets of
classical trajectories lead to the interference effects which cause a hard gap
in the density of states.  This treatment (\cf the reservations in \ocite{vl03})
builds on the recent advances in identifying~\cite{rs02},
codifying~\cite{heusleretal06,mulleretal07} and generating~\cite{bhn08} the
semiclassical contributions, and, because of the slow convergence of the
expansion for the density of states in \eref{dossemieqn}, relies on the ability
to treat correlations between $n$ trajectories for essentially all $n$.  The
correlations between these trajectories, encoded in encounter regions where they
differ slightly, are represented by simple (tree) diagrams.  These diagrams are
related to those that appear for the conductance~\cite{heusleretal06} say where
for increasing $n$ they cause ever decreasing (in inverse channel number)
corrections; here though they all contribute with roughly the same (slowly
decreasing) importance.  It is because we need to treat all orders that
Andreev billiards are so interesting and the resultant effects so large.

Along with obtaining the minigap, found by RMT~\cite{melsenetal96}, for a
billiard with a single lead, we could also obtain the full result for the
density of states of an Andreev billiard with two superconducting leads at phase
difference $\phi$, treated using RMT in \ocite{melsenetal97}.  The semiclassical
paths that connect the two leads accumulate phases $\rme^{\pm\rmi\phi}$ and
cause the gap to shrink with increasing phase difference.  It was also possible
to treat the effect of a time reversal symmetry breaking magnetic field $b$,
considered with RMT in \ocite{melsenetal97}, which makes the formation of the
classical trajectory sets, traversed in opposite directions by an electron and a
hole, less likely.  This in turn leads to a reduction of the minigap and a
smoothing of the density of states, especially for large phase differences
$\phi$.  We have found that in the limits $\phi\to\pi$ and $b\to\infty$ quantum
effects vanish and the density of states becomes identical to the density of
states of the isolated billiard.

The agreement shown here between the semiclassical and the RMT results may lead
one to wonder about the deeper connections between the two methods.  Indeed the
diagrammatic methods~\cite{bb97} used in RMT, which also use recursion relations
over planar diagrams, bear some resemblance to the tree recursions here, but
unfortunately any correspondence between the two is somewhat obscured by the
transformation from the generating function $G$ to the generating function $H$. 
It is also worth pointing out that our semiclassical treatment (with its inherent
semiclassical limits) gives us access to the typical and universal density of
states of chaotic systems.  However there can be further effects that change
the shape of the density of states, for example scarring~\cite{ks06}, classical
Lyapunov exponent fluctuations~\cite{silvestrov06} and
disorder~\cite{libischetal08}.

Of course all our results in this paper (and the RMT
ones~\cite{melsenetal96,melsenetal97}) are only valid to leading order in
inverse channel number.  With the formalism shown in this paper, to go to
subleading order we only require a way of generating the possible semiclassical
diagrams.  The contribution of each~\cite{heusleretal06,mulleretal07} and how
they affect the density of states is known in principle, but the key problem is
that the structure we used here breaks down, namely that in the tree recursions
when we cut a rooted plane tree at a node we created additional rooted plane
trees~\cite{bhn08}.  How to treat the possible diagrams which include closed
loops etc, though generated for $n=1$ in \ocite{heusleretal06} and $n=2$ in
\ocite{mulleretal07} by cutting open closed periodic orbits, remains unclear. 
However the treatment for $n=1$ and $n=2$ makes it clear that the diagrams that
contribute at order $(1/N^{m},n)$ are related to those that contribute at order
$(1/N^{m-1},n+1)$, raising the possibility of a recursive treatment starting from
the leading order diagrams described here.

Worth noting is that the semiclassical techniques we used here are only valid up
to the Heisenberg time, meaning that we have no access to the density of states
on energy scales of the order of the mean level spacing.  Though for ballistic
transport the Heisenberg time is much longer than the average dwell time (so the
mean level spacing is much smaller than the Thouless energy) importantly the RMT
treatment~\cite{frahmetal96} shows that a microscopic gap persists in this
regime even when the time reversal symmetry is completely broken (by the
magnetic field say).  It may be possible that applying the semiclassical
treatment of times longer than the Heisenberg time for closed
systems~\cite{heusleretal07,mulleretal09} to transport would allow one to access
this regime as well.

In the opposite regime however, that of the Ehrenfest time, semiclassics
provides a surprisingly simple result~\cite{waltneretal10} allowing complete
access to the crossover from the universal RMT regime to the more classical
Bohr-Sommerfeld regime.  The gap shrinks due to the suppression of the formation
of encounters while a new class of diagrams (correlated bands) becomes possible. 
Interestingly the contribution from trajectories with encounters agrees exactly
with the results from effective RMT~\cite{beenakker05}, so our semiclassical
result provides support for this phenomenological approach.  Of course effective
RMT misses the bands of correlated trajectories (\cf those in \ocite{ma09})
which combined with the other contribution lead to new effects, most notably a
second gap in the density of states for intermediate Ehrenfest times.

\begin{acknowledgments}
The authors would like to thank \.{I}.~Adagideli for useful conversations and
gratefully acknowledge the Deutsche Forschungsgemeinschaft within GRK 638 (DW,
KR) and FOR 760 (KR), the National Science Foundation under grant 0604859 (GB), CEA Eurotalent (CP)
and the Alexander von Humboldt Foundation (JK, CP) for funding.
\end{acknowledgments}

\begin{widetext}

\appendix*

\section{Further generating functions}

The intermediate generating function $I(\en,r)$ for the billiard with a single
lead and no magnetic field in \sref{singledos} is given by
\begin{eqnarray}
\nonumber && 1- \left[ \left(1-a\right)^{2}+6r+\left(1+a\right)^{2}r^2 \right] I
+ \left[
4\left(1-a\right)^{3}-\left(8+20a^2-a^4\right)r+4\left(1+a\right)^{3}r^2 \right]
rI^{2}\\
\nonumber && {} + \left[
4\left(1-a\right)^{3}
-\left(16-24a+44a^2-8a^3-a^4\right)r+2\left(12+32a^2-a^4\right)r^2 \right. \\
&& \left. \qquad {}
-\left(16+24a+44a^2+8a^3-a^4\right)r^3+4\left(1+a\right)^{3}r^4 \right] rI^{3} =
0 ,
\label{simpleIeqn}
\end{eqnarray}
where we set $a=\rmi\en$.

The generating function $H(\en,\phi,b,r)$ for the billiard with equal leads at
phase difference $\phi$ and magnetic field $b$ in \sref{twomag} is given by
\begin{eqnarray}
\nonumber && {\beta}^{2}- \left(  \left( 1-a+b \right) ^{2}+{r}^{2}-2 r \left(
1-a
+b \right)  \left( 2 {\beta}^{2}-1 \right)  \right) H \\
\nonumber && {} -r \left[\left( 1-a+b \right)  \left( 1-3 a+7 b \right) -2 r
\left( 1+5 b+
{b}^{2}- \left( 3+2 b \right) a+{a}^{2} \right)  \left( 2 {\beta}^{2
}-1 \right)+{r}^{2} \left( 1-2 a+2 b \right)  \right] {H}^{2}\\
\nonumber && {}+{r}^{2} \left[ -b \left( 19 b+10 \right) +2 a \left( 9 b+1
\right) -3 {
a}^{2} +2 r \left( 2 b \left( 3 b+4 \right) -2 a \left( 4 b+1
 \right) +2 {a}^{2} \right)  \left( 2 {\beta}^{2}-1 \right) \right. \\
\nonumber && \left. \qquad {} +{r}^{2}
 \left( -b \left( b+6 \right) +2 a \left( b+1 \right) -{a}^{2}
 \right)  \right] {H}^{3}\\
\nonumber && {} -{r}^{3} \left[ b \left( 25 b+4 \right) -14
 ab+{a}^{2}-2 r \left( b \left( 13 b+4 \right) -10 ab+{a}^{2}
 \right)  \left( 2 {\beta}^{2}-1 \right) +{r}^{2} \left( b \left( 5 
b+4 \right) -6 ab+{a}^{2} \right)  \right] {H}^{4}\\
\nonumber && {} -4 {r}^{4}b \left[ 4 b-a-2 r \left( 3 b-a \right)  \left( 2
{\beta}^{2}-1
 \right) +{r}^{2} \left( 2 b-a \right)  \right] {H}^{5} \\
&& {} -4 {r}^{5}{b}^{2} \left[ 1+{r}^{2}-2 r \left( 2 {\beta}^{2}-1 \right) 
\right] {H
}^{6} =0 ,
\label{magphaseHeqn}
\end{eqnarray}
where we also used $a=\rmi\en$.  For the billiard with unequal leads and no
magnetic field in \sref{unequal}, the generating function $H(\en,\phi,y,r)$ is
given by
\begin{eqnarray}
\nonumber && \beta {\beta^{*}} \left( 1-a \right) ^{2}+\beta {\beta^{*}}
{r}^{2}- \left( {\beta}^{2}+{{\beta^{*}}}^{2} \right)\left( 1-a \right) r \\
\nonumber && {}+ \left[ - \left( 1-a \right) ^{4}+r \left(\left(
\beta+{\beta^{*}} \right) ^{2} \left( 1-{a}^{3} \right) 
+ \left( 3  \left( \beta+{\beta^{*}} \right) ^{2}+2 \beta {\beta^{*}} \right) a
\left( a-1 \right)  \right) \right. \\
\nonumber && \qquad {} +{r}^{2} \left( 
 \left( 3  \left( \beta+{\beta^{*}} \right) ^{2}-2 \beta
 {\beta^{*}}-2 \right) a \left( 2-a \right) +2  \left( 1+\beta
+{\beta^{*}} \right)  \left( 1-\beta-{\beta^{*}} \right) 
 \right) \\
\nonumber && \qquad \left. {} +{r}^{3} \left(  \left( \beta+{\beta^{*}} \right)
^{2}-
a \left(  \left( \beta+{\beta^{*}} \right) ^{2}+2 \beta 
{\beta^{*}} \right)  \right) -{r}^{4} \right] H \\
\nonumber && {} +r \left[  \left( 1-a \right) ^{3} \left( 5 a-1 \right) + \left(
 \left( \beta+{\beta^{*}} \right) ^{2} \left( 1-7 a-7 {a}^{3}+{a}^{4} \right) +
\left( 
3 \beta+4 {\beta^{*}} \right)  \left( 4 \beta+3 {\beta^{*}} \right) {a}^{2}
\right) r \right.\\ 
\nonumber && \qquad {}+  2  \left( 1+\beta+{\beta^{*}} \right)  \left(
1-\beta-{\beta^{*}} \right)  \left( 1
-6 a-2 {a}^{3} \right)r^2 - \left( 15 {\beta}^{2}+15 {{\beta^{*}}}^{2}-14+28
\beta {\beta^{*}} \right) {a}^{2} {r
}^{2} \\
\nonumber && \left. \qquad {} + \left(  \left( \beta+{\beta^{*}} \right) ^{2}
\left( 1-5
 a \right) + \left( 3 {\beta}^{2}+3 {{\beta^{*}}}^{2}+7 {
\beta} {\beta^{*}} \right) {a}^{2} \right) {r}^{3}+ \left( 4 a-1
 \right) {r}^{4} \right] {H}^{2} \\
\nonumber && {} +a{r}^{2} \left[ 2  \left( 1-a \right) ^{2} \left( 2-5 a \right)
+ \left( \beta+{\beta^{*}}
 \right) ^{2} \left( 4 {a}^{3}-15 {a}^{2}+15 a-4 \right) r \right. \\
\nonumber && \qquad {} +2\left( 1+\beta+{\beta^{*}} \right)  \left(
1-\beta-{\beta^{*}} \right)  \left( {a}^{3}-8 {a}^{2}+12 a-4 \right) {r}^{2} \\ 
\nonumber && \left. \qquad {}+\left( \beta+{\beta^{*}} \right) ^{2} \left( -3
{a}^{2}+9 a-4
 \right) {r}^{3}+ \left( 4-6 a \right) {r}^{4} \right] {H}^{3} \\
\nonumber && {}+{a}^{2}{r}^{3} \left[ 16 a-10 {a}^{2}-6+ \left(
\beta+{\beta^{*}}
 \right) ^{2} \left( 6-13 a+6 {a}^{2} \right) r +2  \left( 1+{\beta}+{\beta^{*}}
\right)  \left( 1-\beta-{\beta^{*}} \right) 
 \left( 6-10 a+3 {a}^{2} \right) {r}^{2} \right. \\
\nonumber && \left. \qquad {}+ \left( \beta+{\beta^{*}} \right) ^{2} \left( 6-7
a+{a}^{2} \right) {r}^{3}+ \left( 4 a
-6 \right) {r}^{4} \right] {H}^{4} \\
\nonumber && {}+{a}^{3}{r}^{4} \left[ 4-5 a+4 \left( \beta+{\beta^{*}} \right)
^{2} \left( a-1 \right) r+2 
 \left( 1+\beta+{\beta^{*}} \right)  \left( 1-\beta-{\beta^{*}} \right)  \left(
3 a-4 \right) {r}^{2} \right.\\ 
\nonumber && \qquad \left.{} + \left( \beta+{\beta^{*}} \right) ^{2} \left( 2
a-4 \right) {r}^{3}+ \left( 4-a
 \right) {r}^{4} \right] {H}^{5} \\
&& {}+{a}^{4}{r}^{5} \left( -1-{r}^{4}+r \left( 1+{r}^{2} \right)  \left(
\beta+{\beta^{*}} \right) ^{2}
+2 {r}^{2} \left[ 1+\beta+{\beta^{*}} \right)  \left( 1-\beta-{\beta^{*}}
\right)  \right] {H}^{6} =0,
\label{unequalphaseHeqn}
\end{eqnarray}
likewise with $a=\rmi\en$.

\end{widetext}

\end{document}